\documentclass[fleqn,usenatbib]{mnras}%

\usepackage{newtxtext,newtxmath}

\usepackage[T1]{fontenc}

\DeclareRobustCommand{\VAN}[3]{#2}
\let\VANthebibliography\thebibliography
\def\thebibliography{\DeclareRobustCommand{\VAN}[3]{##3}\VANthebibliography}

\usepackage{graphicx}	
\usepackage{amsmath}	
\usepackage{subfig}
\usepackage{color, colortbl}
\usepackage{soul}

\def \aap{A\&A }
\def \apjl{ApJ}
\def \apjs{ApJS}
\def \apj{ApJ}
\def \jcap{JCAP}

\def \mnras{MNRAS}

\def \prl{Phys.~Rev.~Lett.}

\def \nat{Nature\ }

\title[CTA sensitivity to a new MSPs population at the GC]{Cherenkov Telescope Array sensitivity to the putative millisecond pulsar population responsible for the Galactic center excess }

\author[Macias et al.]{
\parbox[t]{\textwidth}{
Oscar Macias,$^{1,2}$\thanks{o.a.maciasramirez@uva.nl}
Harm van Leijen,$^{2}$
Deheng Song,$^{3}$
Shin'ichiro Ando,$^{2,1}$
Shunsaku Horiuchi,$^{3}$ and 
Roland M. Crocker$^{4}$\\}
\\
$^{1}$Kavli Institute for the Physics and Mathematics of the Universe (WPI), University of Tokyo, Kashiwa, Chiba 277-8583, Japan\\
$^{2}$GRAPPA Institute, Institute of Physics, University of Amsterdam, 1098 XH Amsterdam, The Netherlands\\
$^{3}$Center for Neutrino Physics, Department of Physics, Virginia Tech, Blacksburg, VA 24061, USA\\
$^{4}$Research School of Astronomy and Astrophysics, Australian National University, Canberra, ACT 2611, Australia\\
}

\date{Accepted XXX. Received YYY; in original form ZZZ}

\pubyear{2021}

\begin{document}
\label{firstpage}
\pagerange{\pageref{firstpage}--\pageref{lastpage}}

\maketitle

\begin{abstract}
The leading explanation of the \textit{Fermi} Galactic center $\gamma$-ray excess is the extended emission from a unresolved population of millisecond
pulsars (MSPs) in the Galactic bulge. Such a population would, along with the prompt $\gamma$ rays, also inject large quantities of electrons/positrons ($e^\pm$) into the interstellar medium. These $e^\pm$ could potentially inverse-Compton (IC) scatter ambient photons into $\gamma$ rays that fall within the sensitivity range of the upcoming Cherenkov Telescope Array (CTA). In this article, we examine the detection potential of CTA to this signature by making a realistic estimation of the systematic uncertainties on the Galactic diffuse emission model at TeV-scale $\gamma$-ray energies. We forecast that, in the event that $e^\pm$ injection spectra are harder than $E^{-2}$, CTA has the potential to robustly discover the IC signature of 
a putative Galactic bulge MSP population sufficient to explain the GCE for $e^\pm$ injection efficiencies in the range $\approx 2.9$--74.1\%, or higher, depending on the level of mismodeling of the Galactic diffuse emission components. On the other hand, for spectra softer than $E^{-2.5}$, a reliable CTA detection would require an unphysically large $e^\pm$ injection efficiency of $\gtrsim 158\%$. However, even this pessimistic conclusion may be avoided in the plausible event that
MSP observational and/or modeling uncertainties can be reduced. We further find that, in the event that an IC signal were detected, CTA can successfully discriminate between an MSP and a dark matter origin for the radiating $e^\pm$.
\end{abstract}


\section{Introduction}
\label{sec:introduction}

\textit{Fermi} Large Area Telescope (\textit{Fermi}-LAT) observations~\citep{Hooper:2010mq,Abazajian:2010zy,Abazajian:2012pn,Gordon:2013vta, Macias:2013vya,Calore:2014xka,Daylan:2014rsa,TheFermi-LAT:2015kwa,TheFermi-LAT:2017vmf} of the inner $\sim 10^{\circ}$ of the Galactic center (GC) region have revealed extended  $\gamma$-ray emission in significant excess with respect to the astrophysical background model. Early studies of this `Galactic center excess' (GCE) appeared to show the signal to be spherically symmetric around the GC with a radially-declining intensity. This, together with the fact that the GCE is described by a 
spectral energy distribution peaked at a few GeV,
led to the possibility that it
originates from
self-annihilation of  weakly interacting massive particles (WIMPs) spatially distributed 
according to something approaching (actually somewhat steeper than)
a Navarro-Frenk-White (NFW) density profile 
\citep{Hooper:2010mq,Abazajian:2012pn,Gordon:2013vta}. However, recent reanalyses of the spatial morphology of the GCE~\citep{Macias:2016nev,Bartels:2017vsx,Macias:2019omb, Abazajian:2020tww} have demonstrated that its spatial morphology is better described by the (non-spherically symmetric) stellar density distribution of the Galactic bulge than by profiles expected for the WIMP annihilation. The Galactic bulge hosts a large variety of stellar populations from old to 
recently-formed
\citep{Garzon:1997cs,Hammersley:2000mx}, and should contain many 
types of
$\gamma$-ray sources, 
the prime example of which is the expected large population of millisecond pulsars (MSPs)~\citep{Abazajian:2010zy,Abazajian:2012pn,Gordon:2013vta,Ploeg:2017vai, Fragione:2017rsp, Fragione:2018jxd, Gonthier:2018ymi, Ploeg:2020jeh}---old pulsars that have been spun-up due to their
interaction within a binary system.

In particular, \citet{Abazajian:2020tww} performed a reanalysis of the GCE and demonstrated that, when including maps tracing stellar mass in the Galactic and nuclear bulges, the GC shows no significant detection of a dark matter (DM) annihilation signal. These results were shown to be robust to generous variations of the astrophysical backgrounds (e.g., combinations of two-dimensional or three-dimensional inverse Compton maps, interstellar gas maps, a central source of electrons) and DM morphologies (e.g., cored DM profiles, generalized NFW, and ellipsoidal versions of these). This allowed some of us to obtain very strong constraints on DM properties. Interestingly, the key result that the spatial morphology of the GCE is better described by a stellar bulge template than by a NFW model has been recently confirmed by an updated analysis with SkyFACT~\citep{Calore:2021bty}. Though we notice that~\cite{DiMauro:2021raz} obtained otherwise.

Several recent studies~\cite[e.g.,][]{Lee:2015fea,Bartels:2015aea, Leane:2019xiy,Balaji:2018rwz,Chang:2019ars,Leane:2020pfc,Leane:2020nmi,Buschmann:2020adf,Calore:2021bty} have also proposed different methods to resolve the source nature of the GCE. Although there is no consensus yet in the community regarding the main results of such methods~\cite[e.g.,][]{Leane:2020pfc,Leane:2020nmi,Buschmann:2020adf,Bartels:2015aea,Balaji:2018rwz}, it should be noted that these studies are probing the photon-count statistics of the GCE, which is an aspect of the signal that is independent of the morphological analyses and results in~\citep{Macias:2016nev,Bartels:2017vsx, Macias:2019omb}. Indeed, the spectrum and spatial morphology are the most basic characteristics of the $\gamma$-ray sources in the sky~\citep{Fermi-LAT:4FGL}, and these two characteristics of the GCE are in strong agreement with a Galactic bulge explanation of the GCE~\cite[e.g.,][]{Abazajian:2020tww}. 

Pulsars have long been predicted to be sources of electron-positron ($e^{\pm}$) pairs~\citep{Erber:1966,Sturrock:1970,Sturrock:1971, Aharonian:1995, Atoyan:1995}. Highly energetic $e^{\pm}$ in the wind regions or inner magnetospheres of pulsars, 
also when released into the general interstellar medium,
can inverse-Compton (IC) scatter ambient photons to very high energies. 
The recent HAWC~\citep{Abeysekara:2017old} observations of extended TeV-scale $\gamma$-ray emission around Geminga and Monogem have provided indirect evidence for TeV $e^{\pm}$ acceleration in normal pulsar nebulae.  Since a multitude of normal pulsars have been observed in the vicinity of the solar system, they have been purported~\citep{Hooper:2008kg, Delahaye:2010,Abeysekara:2017old,Hooper:2017gtd, Profumo:2018fmz, DiMauro:2019yvh, Johannesson:2019jlk} as the likely source of the majority of local cosmic ray positrons~\citep{Aguilar:2013qda}.    
Several lines of evidence also point to potentially efficient $e^{\pm}$ acceleration by MSPs. First, a recent study~\citep{Sudoh:2020hyu} of the correlation between far-infrared and radio luminosities in star-forming galaxies (SFGs) found that radio emission from MSPs may account for a large fraction of the radio luminosity observed in systems with high stellar mass but low star formation rate. 
In particular, the unexpectedly high radio luminosities of such
systems
might be explained by the cooling of GeV-scale $e^{\pm}$ from MSPs via synchrotron radiation in each host galaxy's interstellar medium (ISM). Second, ~\citet{Song:2021zrs} have found evidence for IC emission produced by MSP populations in globular clusters of the Milky Way. Furthermore, the H.E.S.S. telescope has observed extended TeV $\gamma$-ray emission from the direction of the globular cluster Terzan 5~\citep{Abramowski:2013md}. Such $\gamma$-rays could naturally arise from Comptonization of the background radiation within the globular cluster by super-TeV $e^{\pm}$~\citep{Bednarek:2016gpp}. 
Note that while a similar search conducted by H.E.S.S in another 15 globular clusters~\citep{Abramowski:2013md}, and MAGIC in the M15 globular cluster~\citep{Acciari:2019ysf}, only produced $\gamma$-ray flux upper limits, it seems likely that forthcoming $\gamma$-ray instruments could resolve a large number of globular clusters at TeV energies~\citep{Ndiyavala:2017hoh,Ndiyavala-Davids:2020wjc}. 
Third, several analyses of the GCE~\citep[e.g.,][]{Horiuchi:2016zwu,Linden:2016rcf,DiMauro:2021raz} have found evidence for a high-energy tail, possibly related to IC emission from $e^\pm$ accelerated by the sources responsible for it. 
Finally, recent simulation work \citep{Guepin:2019fjb} suggests that MSPs could efficiently accelerate $e^{\pm}$ pairs to TeV-scale energies and beyond. 

In this paper, we propose that the hypothesized population of approximately 20 to 50 thousand MSPs~\citep{Ploeg:2020jeh} in the Galactic bulge, could inject large numbers of highly energetic $e^{\pm}$ into the ISM~\footnote{Note that other studies~\citep[e.g.,][]{Guepin:2018jkb} have explored the proton signatures from GC MSPs.}. The  IC signal resulting from these could be detectable~\citep{Song:2019nrx} by planned TeV-scale $\gamma$-ray telescopes such as 
the Cherenkov Telescope Array (CTA). 
In particular, CTA will observe the center of the Milky Way with unprecedented spectral and spatial detail~\citep{CTAConsortium:2018tzg}. 
CTA's anticipated strategy for a survey of the inner Galaxy entails a deep exposure observation of the inner few degrees of the GC region plus an extended survey covering a large fraction of the northern side of the Galactic bulge ($|l|\lesssim 6^{\circ}\;{\rm and}\; 0.3^{\circ}\lesssim b \lesssim 10^{\circ}$). The latter will facilitate the study of diffuse very high energy (VHE) sources such as Galactic outflows (a possible VHE counterpart to the Fermi bubbles), interstellar gas-correlated $\gamma$-ray emission, a potential DM emission signature~\citep{Acharyya:2020sbj}, unresolved $\gamma$-ray sources~\citep{Viana:2019jwd}, and, as we will thoroughly explore here, the IC emission from the 
population of $e^\pm$ launched into the bulge ISM by the
 MSPs putatively responsible for the GCE.

 Recently, ~\cite{Song:2019nrx} performed detailed simulations (using \textsc{galprop} V54~\citealt{Porter:2006tb}) of the IC signature from the putative population of MSPs responsible for the GCE. Compared to that work, here we recompute all our spatial maps using the latest version of the code (\textsc{galprop} V56)---which contains new three-dimensional (3D) models for the interstellar radiation fields, Galactic structures, and interstellar gas maps. Furthermore, now we run our simulations in the 3D mode of the \textsc{galprop} framework, abandoning the assumption of Galactocentric cylindrical symmetry. Interestingly, \cite{Song:2019nrx} argued that it could be possible for CTA to (i) detect the population of MSPs responsible for the GCE, and (ii) distinguish whether the IC signal emanates from either DM self-annihilation or these unresolved MSPs in the Galactic bulge. 
 
 Here, we perform a realistic assessment of the sensitivity of CTA to such an unresolved population of MSPs in the GC. Using simulated data, we consider scenarios where the Galactic diffuse emission (GDE) is known accurately, as well as cases where the GDE is mismodeled. In each case, we investigate the necessary conditions for a reliable CTA detection of the MSPs' IC signal. Similarly, we study whether CTA could separate the nature of the source producing the IC emission. As shown in~\citet{Song:2019nrx}, the expected IC emission from the Galactic bulge MSPs has morphological differences with respect to the one from DM emission. We now present a systematic study on simulated data that takes into account the latest CTA instrument response function and state-of-the-art models for the GDE to fully address these points.
 
This paper is organized as follows. In Sec. \ref{sec:ICfromMSPs} we describe the procedure used to construct the TeV-scale IC flux maps produced by the putative MSP population responsible for the GCE. In Sec.~\ref{sec:backgroundtemplates}, we present the astrophysical backgrounds in the GC region that are relevant for the IC searches. In Sec.~\ref{sec:sensitivity} we provide an overview of the expected CTA performance, the expected background and signal rates, and the methodology to estimate the CTA sensitivity. We present our results in Sec.~\ref{sec:results}, and the discussion and conclusions in Sec.~\ref{sec:discussions}.

\section{Computation of IC emission from MSPs in the Galactic Center}
\label{sec:ICfromMSPs}

\subsection{Production mechanisms of $e^{\pm}$ pairs in millisecond pulsars}
\label{subsec:e+-injection}

Millisecond pulsars are neutron stars with high rotation rates (spin periods in the range of tens of ms or less) and relatively low surface magnetic fields ($\sim 10^{8}$ G). Due to their very rapid rotation, MSPs can produce high  
electric fields and spin-down power. Despite their low surface magnetic fields, their very compact magnetospheres allow for 
magnetic fields at the light cylinder that are comparable to those of normal pulsars~\citep{Harding:2021yuv}. Their observed pulsed emission---as seen by an observer located inside the cone scanned by the magnetic field---is explained by the misalignment of the pulsar rotation axis and the magnetic field axis.  If the neutron star surface temperature ($T_s$) is greater than the electron surface binding temperature (i.e., $T_s\gtrsim 10^5$ K), then electrons can be torn from the neutron star surface by the strong electric fields~\citep{Michel:1991}. Such electrons would subsequently move parallel to magnetic field lines and emit $\gamma$-ray photons through curvature radiation. An additional large number of ``secondary'' electrons (and positrons) can be created through pair production processes at different sites throughout the MSPs magnetosphere. 
Dedicated phenomenological studies of MSPs demonstrate that most of their high-energy emission occurs near the current sheet outside the light cylinder~\citep{Fermi-LAT:4FGL,Harding:2021yuv}
Furthermore, in global magnetosphere models, charge currents, electromagnetic fields, and current sheets scale with the light cylinder independent of the pulsar surface magnetic field strength or rotation period. Thus, new emission models of MSPs predict high energy radiation that is similar to that of normal pulsars (the reader is referred to~\citealt{Harding:2021yuv} for further details).

\subsection{Spatial distribution of the MSP population in the Galactic Center}
\label{subsec:spatialdistribution}

A very interesting implication of recent GCE analyses~\citep{Macias:2016nev,Bartels:2017vsx,Macias:2019omb,Abazajian:2020tww}, is that prompt $\gamma$-ray emission from MSPs in the Galactic bulge could account for the bulk of the GCE emission~\citep{Abazajian:2012pn}. Since 
prompt emission occurs within or very close to the magnetospheres of individual MSPs and
$\gamma$ rays travel following geodesics in space-time, the spatial morphology of the GCE can be used to map the spatial distribution of this putative MSP population.

Here, we follow the same approach as \cite{Song:2019nrx} who assumed that the MSP population is smoothly distributed following the density of stars in the Galactic bulge (see Sec.~\ref{subsec:disentanglingMSPsfromDM}). 
In particular, the Galactic bulge consists of two main components: the nuclear bulge, and the boxy bulge (see Appendix~\ref{appdx:Stellardensity}). The former corresponds to the stellar structures residing in the inner $\sim 200$ pc of the Galactic center (the nuclear stellar disk; and the nuclear star cluster~\citealt{Launhardt:2002tx}), and the latter refers to the stars residing in the inner $\sim3$ kpc of the Galactic bar~\citep{Freudenreich:1998,Coleman:2019kax}. However, we also consider the 
case where the $e^\pm$ sources are distributed with spherical symmetry as this would be spectrally and spatially similar to the IC signal from DM annihilation. 
More specifically, for the spherical distribution of 
$e^\pm$ sources we use a generalized Navarro-Frenk-White (NFW) profile with a mild radial slope. 
Detailed descriptions of the stellar density maps, and the NFW map used in this work are given in Appendix~\ref{appdx:Stellardensity}. 

A potentially relevant effect, which our simulations do not consider, corresponds to the possible smearing effect on the source distribution caused by the MSPs kick velocity distribution. However, it is expected that the kicks experienced by MSPs are lower ($\approx 10$--50~km~s$^{-1}$) than for normal pulsars~\citep{Podsiadlowski:2005}. This comes mainly from the observation that the escape velocity in globular clusters ($\lesssim 50$~km~s$^{-1}$, according to~\citealt{Pfahl:2001df}) is much lower than the average kick velocity of normal pulsars ($\sim 250$~km~s$^{-1}$;~\citealt{Hobbs:2005yx,Atri:2019fbx}), yet globular clusters have been observed to contain a large number of MSPs. For initial kick velocities $\lesssim 20$~km~s$^{-1}$, we estimate a negligible smearing of the Galactic bulge stellar density maps. However, we will make a more in-depth investigation of the impact of this effect in a future study of this subject. 

\subsection{Propagation of CRs in the Galaxy}
\label{subsec:propagation}

Given the aforementioned spatial distribution of MSPs in the Galactic bulge, \cite{Song:2019nrx} thoroughly studied the injection and propagation of highly energetic $e^{\pm}$ from MSPs into the interstellar medium. The focus of that study was the construction of high-resolution IC maps for TeV-scale morphological analyses of future $\gamma$-ray observations. There, it was shown that while the prompt $\gamma$ rays from MSPs trace the $\gamma$-ray source distribution, their IC counterpart exhibits an energy-dependent spatial morphology. This is due to the IC emission being the result of 
the convolution of the relevant photon targets -- starlight, infrared (IR) light, the cosmic microwave background (CMB) -- with the
steady state cosmic-ray (CR) distribution, which has itself 
a spatially-dependent spectrum because it is, in turn, the product of the CR source distribution and energy-dependent loss and diffusion processes.

Similar to the approach of~\citet{Song:2019nrx} and \citet{Ishiwata:2019aet}, here, we solve the CRs transport equation for MSP-accelerated $e^{\pm}$ using a customized version of the numerical propagation code \textsc{galprop}~\citep{Porter:2006tb,Strong:2007nh}. Given a certain CR source distribution, CR injection energy spectrum, and interstellar medium properties, \textsc{galprop} makes self-consistent predictions  
for the spatial distributions and spatially-dependent spectra of
 all CR species. The \textsc{galprop} framework includes pure diffusion, convection, diffusive re-acceleration, and energy losses (e.g., in the lepton case: Coulomb scattering, synchrotron, inverse Compton scattering, bremmstrahlung,  and nuclear collisions).

\citet{Song:2019nrx} computed the propagation of $e^{\pm}$ with \textsc{galprop} version 54 (v54). This version of the code makes some simplifying assumptions for the propagation of Galactic CRs. Namely, it  assumes galactocentric cylindrical symmetry of the CR halo, two-dimensional (2D) interstellar radiation fields (ISRF) model, and 2D interstellar gas maps. Importantly, it has been pointed out~\citep{Porter:2017vaa,Johannesson:2018bit} that these assumptions could significantly impact the predictions of IC, and gas-correlated $\gamma$-ray emission. In view of this fact, for the present article, we update the IC calculations of \citet{Song:2019nrx} by using the latest release of the code [\textsc{galprop} v56~\citep{Porter:2017vaa,Johannesson:2018bit}].    

Unlike previous versions of this numerical code, \textsc{galprop v56}  contains more sophisticated 3D models of the ISRF, 
neutral hydrogen, and molecular hydrogen gas. Such recent improvements of the code have allowed for detailed predictions of anisotropies~\citep{Porter:2017vaa,Johannesson:2018bit} in the Galactic diffuse $\gamma$-ray emission and better, physically-motivated solutions for the CR transport equation.
 Currently, there are two alternative 3D ISRF models that are available to the user: the first one is based on the Galaxy model proposed by~\cite{Freudenreich:1998} (F98), and the second one is based on~\cite{R12} (R12). Despite that these two models assume different dust densities; and stellar luminosities, their predicted local fluxes are both consistent with near/far-infrared measurements. Motivated by previous analyses of the GCE~\citep{Macias:2016nev,Bartels:2017vsx,Macias:2019omb}, we limit our study to the F98 Galactic structure model within \textsc{galprop v56}.  

\begin{table}
\centering
\begin{tabular}{lc}\hline\hline
Parameter                                              & value\\\hline
$X_h$ [kpc]                                            & $\pm 20.00$\\
$Y_h$ [kpc]                                            & $\pm 20.00$\\
$Z_h$ [kpc]                                            & $\pm 6.00$\\
$\Delta X$ [kpc]                                       & $0.2$\\
$\Delta Y$ [kpc]                                       & $0.2$\\
$\Delta Z$ [kpc]                                       & $0.1$\\
$D_{0,xx}$ [$10^{28}$ cm$^2$s$^{-1}$]                  & $2.28$\\
$\delta$                              & $0.545$\\
$V_{\rm Alfven}$ [km s$^{-1}$]                         & $5.26$ \\
$\gamma_0$                            & $1.51$ \\
$\gamma_1$                            & $2.35$\\
$R_1$ [GV]                            & $3.56$\\
$\gamma_{0,H}$                        & $1.71$\\
$\gamma_{1,H}$                        & $2.35$\\
$\gamma_{2,H}$                        & $2.19$\\
$R_{1,H}$ [GV]                        & $4.81$\\
$R_{2,H}$ [GV]                        & $200$\\
$\gamma_{0,e}$                        & $1.81$\\
$\gamma_{1,e}$                        & $2.77$\\
$\gamma_{2,e}$                        & $2.38$\\
$R_{1,e}$ [GV]                        & $5.97$\\
$R_{2,e}$ [GV]                        & $76$\\\hline\hline
\end{tabular}\caption{\textsc{galprop} propagation parameter setup. The parameters $(X_h, Y_h, Z_h)$, and $(\Delta X, \Delta Y, \Delta Z)$, give the size of the CR halo; and the spatial resolution (grid size) in Cartesian coordinates, respectively. The diffusion coefficient is a function of rigidity (i.e., $D(R) \propto \beta R^{\delta}$). The injection for CR protons is given by $Q(R) \propto R^{\gamma_{0,H}}$ for $R < R_{1,H}$, $Q(R) \propto R^{\gamma_{1,H}}$ for $R_{1,H} < R < R_{2,H}$, and $Q(R) \propto R^{\gamma_{2,H}}$ for $R > R_{2,H}$. Similarly, for CR $e^-$s it is given by $Q(R) \propto R^{\gamma_{0,e}}$ for $R < R_{1,e}$, $Q(R) \propto R^{\gamma_{1,e}}$ for $R_{1,e} < R < R_{2,e}$, and $Q(R) \propto R^{\gamma_{2,e}}$ for $R > R_{2,e}$. The injection spectrum of heavier  CR nuclei is written as $Q(R) \propto R^{\gamma_0}$ for $R < R_1$, $Q(R) \propto R^{\gamma_1}$ for $ R > R_1$. Other parameters included here are the Alfven velocity ($V_{\rm Alfven}$). See~\citet{Johannesson:2018bit} for more details.   
}\label{tab:galpropsetup}
\end{table}

In the present work, we assume a steady-state solution for the CR transport equation, a homogeneous and isotropic diffusion coefficient, allow for diffusive reacceleration,  and neglect advection of $e^\pm$ in the Galaxy. All the simulations included in our analysis are
performed with the 3D mode of \textsc{galprop}. We also adopt the CR source density model labeled as SA50 in~\cite{Johannesson:2018bit}.  A brief summary of the main parameter values selected for our analysis are shown in Table~\ref{tab:galpropsetup}. \cite{Johannesson:2018bit} obtained these propagation parameters following a similar procedure to that presented in~\cite{Porter:2017vaa}. As can be seen in Table~\ref{tab:galpropsetup}, for the 3D spatial grid we use a bin size of $200\times200\times100$ pc$^3$. Given that nuclear bulge region has a radius of $\approx 200$ pc, this spatial resolution is far from ideal for modeling the diffusion processes in this region. However, 
increasing it further would be very difficult due to very high computing memory demands. We note that our simulations are run in a dedicated computer cluster with 256 GBytes of memory per node. But, since the code only supports \textsc{OpenMP} environments, it is currently not possible to run fully \textsc{MPI} parallelized simulations.  

Given that $e^{\pm}$ could efficiently lose energy through synchrotron radiation, it is important to select well motivated parameters for the random magnetic field of the Galaxy. We use the results of a model that matches the 408 MHz synchrotron data~\citep{Strong:1998fr}, and that agrees with total magnetic field estimates~\citep{1995ASPC...80..507H,2001SSRv...99..243B}. In particular, we use the default double-exponential model included in \textsc{galprop} 

\begin{equation}\label{eq:Bfield}
	B(r,z) = B_0 \exp{\left(-\dfrac{r-R_\odot}{R_0}\right)}\exp{\left(-\dfrac{z}{z_0}\right)},
\end{equation}
where $B_0 = 5\ \mu$G, $R_0 = 10$ kpc, and $z_0 = 2$ kpc. These parameter values are the same as used by~\citet{Johannesson:2018bit}. For the purpose of our current study, we utilize the same random magnetic field parameter setup in all simulations. However, 
we note that \cite{Song:2019nrx} explored other magnetic field configurations~\citep{Crocker:2010xc} to evaluate their impact on the predicted IC maps.

\subsection{Injection Luminosity}

We assume that the Galactic bulge MSPs are injecting $e^{\pm}$ at a constant rate into the interstellar medium with a fraction ($f_{e^{\pm}}$) of their spin-down power ($\dot E$) converted into $e^{\pm}$ pairs. Similarly, the prompt $\gamma$-ray luminosity ($L_{\gamma,{\rm prompt}}$) is assumed to be proportional to the MSPs' spin-down power, whose efficiency $f_\gamma \equiv L_{\gamma,{\rm prompt}}/\dot{E}$, is estimated to be about 10\% on average \citep{TheFermi-LAT:2013ssa}. We can hence write 
\begin{equation}\label{eq:luminosity}
  L_{e^\pm}=f_{e^\pm}\dot{E}=\frac{f_{e^\pm}}{f_{\gamma}}  L_{\gamma,{\rm prompt}} \simeq 10f_{e^\pm} L_{\gamma,{\rm prompt}},
\end{equation}
where $L_{e^\pm}$ is the $e^\pm$ injection luminosity. The $e^\pm$ efficiency $f_{e^\pm}$ has been estimated to be between approximately 7\% and 29\% (computed for $e^\pm$ energies greater than 10 GeV) from TeV observations of normal pulsars~\citep{Hooper:2017gtd}.
Interestingly, a recent phenomenological analysis of MSPs in globular clusters~\citep{Song:2021zrs} finds $f_{e^\pm}\approx 10\%$, while \citet{Sudoh:2020hyu} analyzed radio continuum data from galaxies with low specific star formation rates obtaining a MSPs $f_{e^\pm}$ that can be even greater than $90\%$.

In our study, the $f_{e^\pm}$ is estimated using Eq.~(\ref{eq:luminosity}), simulated GC CTA observations, and the $\gamma$-ray luminosity from \textit{Fermi}-LAT observations of the GC. We use the fit results in \cite{Macias:2019omb}, which for the boxy bulge obtained $L^{\text{bulge}}_{\gamma,{\rm prompt}}$ = (2.2 $\pm$ 0.4) $\times$ 10$^{37}$ erg s$^{-1}$, and for the nuclear bulge $L^{\text{NB}}_{\gamma,{\rm prompt}}$ = (3.9 $\pm$ 0.5) $\times$ 10$^{36}$ erg s$^{-1}$. A fit that instead of the stellar templates included a DM (NFW$^2$) template gave a luminosity estimate of $L^{\text{NFW}^2}_{\gamma,{\rm prompt}}$ = (2.7 $\pm$ 0.4) $\times$ 10$^{37}$ erg s$^{-1}$. These two different morphologies are considered in order to investigate the capabilities of CTA to distinguish the two hypotheses under realistic conditions.

\subsection{Injection spectrum}\label{subsec:spectrum}

For the injection spectrum of the population of GC MSPs we 
assume a power law with an exponential cutoff of the form 
\begin{equation}
  \label{eq:mspspectrum}
  \dfrac{d^2N}{dEdt} \propto E^{-\Gamma}\exp(-E/E_{\text{cut}}),
\end{equation}
where $\Gamma$ is the spectral slope and $E_{\text{cut}}$ is the energy cutoff. While in the case of Supernova Remnants the spectral slope can be constrained from radio measurements, the spectral slope associated with highly energetic $e^\pm$ from pulsars is very difficult to constrain by radio measurements of the pulsed emission~\citep{Delahaye:2010}. This is due to the fact that the pulsed emission is expected to originate in the polar cap region close to the pulsar magnetosphere, whereas a large portion of the most energetic $e^\pm$ pairs are thought to be accelerated by magnetic reconnection in the equatorial current sheet outside the pulsar light cylinder~\citep{Cerutti:2015hvk}. Similarly, the maximum energy that $e^{\pm}$ can attain from MSPs is very uncertain. The $e^\pm$ pairs produced by MSPs are expected to have much higher energies than those of normal pulsars given that MSPs have much lower magnetic fields and therefore the pair-producing photons must have higher energies~\citep{Harding:2021yuv}. Indeed, recent particle-in-cell simulations by~\citep{Guepin:2019fjb} posit that millisecond pulsars could accelerate $e^\pm$ pairs even up to PeV energies.

 Here, we consider several values that have been explored in the literature~\citep[e.g.,][]{Yuan:2014yda,Song:2019nrx,Guepin:2019fjb} and that are expected to encompass the range of uncertainties in the energy cutoff and the spectral slope of the MSPs injection spectrum. In particular, we select several possible parameter values for $\Gamma$, and $E_{\text{cut}}$, as shown in Table~\ref{tab:mspspectrum}. 
 Though each individual MSP will surely have a different age, spin-down luminosity, rotation period, magnetic field, hence a different $e^\pm$ injection spectrum, here we assume that the injection spectrum of the whole population of GC MSPs is characterized by a mean injection spectrum (as shown in Table~\ref{tab:mspspectrum}).
 
\begin{table}
  \centering
    \begin{tabular}{lcc}\hline\hline
    Model Name&$\Gamma$ & $E_{\text{cut}}$\\
    & &  (TeV) \\\hline
    Baseline &2.0 & 50 \\
    Inj1&1.5 & 50 \\
    Inj2&2.5 & 50 \\
    Inj3&2.0 & 10 \\
    Inj4&2.0 & 100\\\hline\hline
    \end{tabular}
    \caption{Injection spectra of $e^{\pm}$ pairs accelerated by a Galactic bulge population of MSPs. See also~\citep{Song:2019nrx}.}
    \label{tab:mspspectrum}
\end{table}

\begin{figure*}
    \begin{center}
    \includegraphics[scale = 0.45]{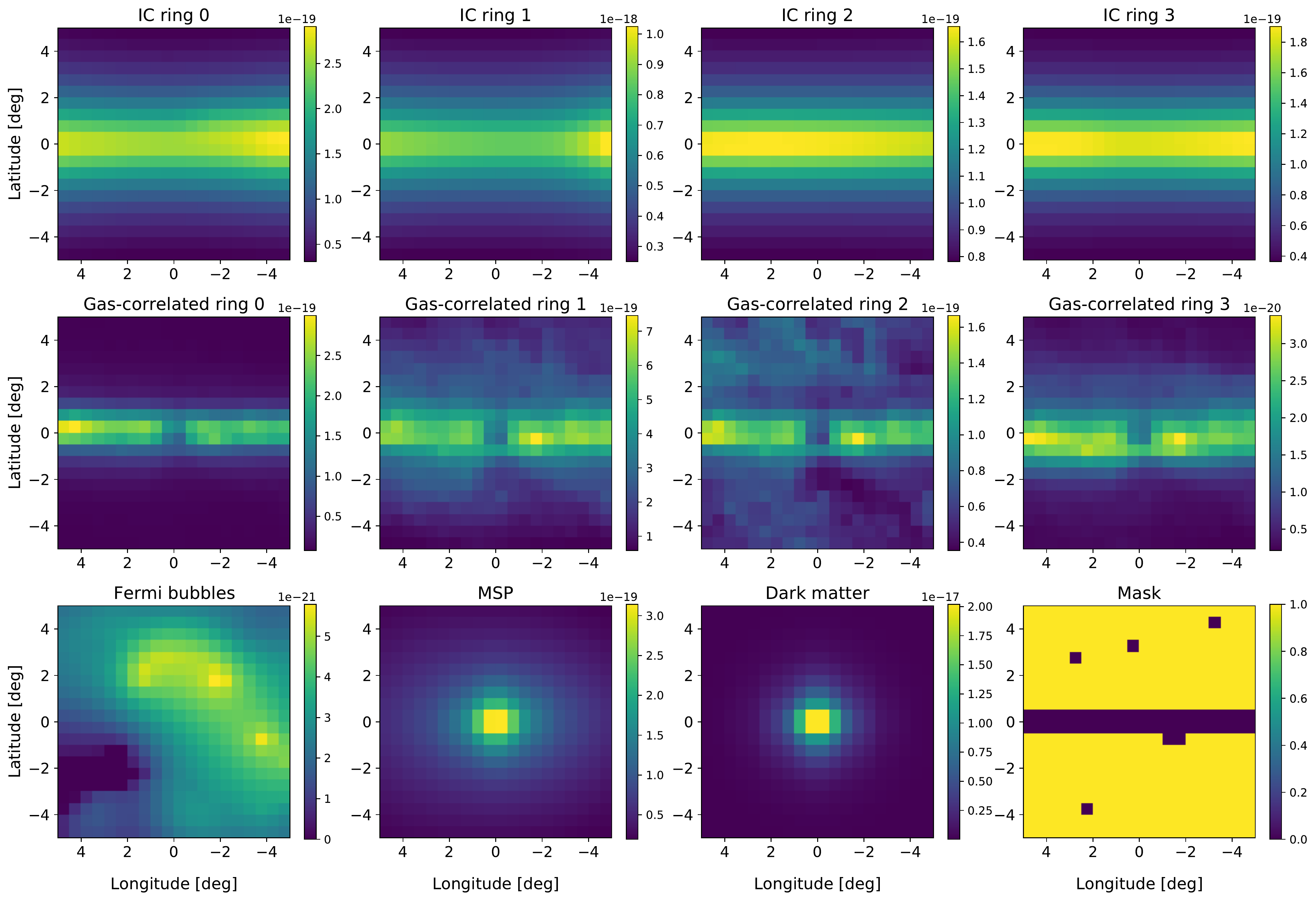}
    \caption{Predicted spatial morphology of various galactic diffusive emission components in the inner $10^\circ \times 10^\circ$ of the Galaxy and at a $\gamma$-ray energy of $\approx 11$ TeV. All maps  
    are shown in arbitrary units, have a spatial resolution of $0.5^\circ \times 0.5^\circ$, and were computed using \textsc{galprop} v56. The top row shows the inverse Compton emission produced by background astrophysical sources, which are divided in four Galactocentric rings (0--3.5, 3.5--8.0, 8.0--10.0, and 10.0--50.0~kpc). The middle row shows the gas-correlated $\gamma$-ray emission maps ($\pi^0+$bremsstrahlung) also divided in rings of the same size. From left to right, the bottom row shows: the Fermi bubbles template introduced in~\citet{Macias:2019omb} (which is based on the one obtained in ~\citet{Fermi-bubbles:2014sfa}), the predicted MSP IC signal at $\approx 11$ TeV, the expected IC signal from a spherical distribution of $e^{\pm}$ sources also at $\approx 11$ TeV 
    (see text for details), and the mask map, respectively. Note that all the maps in this panel (except for the mask, of course) are normalized to \textit{Fermi}-LAT measurements using the procedure explained in the section corresponding to each map (see also Fig.~\ref{fig:counts_spectra}).}
    \label{fig:galpropmaps}
    \end{center}
\end{figure*}

Using Eqs.~(\ref{eq:luminosity}) and (\ref{eq:mspspectrum}), we can obtain the source term $q(\vec{r},E)$ [with units; MeV$^{-1}$ cm$^{-2}$ s$^{-2}$ sr$^{-1}$] in the CR transport equation to be included to our customized version of \textsc{galprop}. Specifically, this can be written as the product of the injection spectrum $d^2N/(dEdt)$, and the MSPs density distribution $\rho(\vec{r})$ as follows 
\begin{equation}\label{eq:sourcefunction}
  q(\vec{r},E) = \frac{c}{4\pi} N_0 \dfrac{d^2N}{dEdt}\rho(\vec{r}),
\end{equation}
where the $c/4\pi$ term is a convention in the \textsc{galprop} code, and $N_0$ is a normalization factor. We normalize the source function in such a way that its integration over volume and energy matches Eq.~(\ref{eq:luminosity}),
\begin{equation}
  \label{eq:norm}
  N_0 \int E\dfrac{d^2N}{dEdt}dE \int\rho(\vec{r})dr^3 = L_{e^\pm}. 
\end{equation}
This procedure is applied for each injection spectrum described in Table~\ref{tab:mspspectrum}, the propagation parameter setup shown in Table~\ref{tab:galpropsetup}, and the spatial distributions given by the stellar mass in the Galactic bulge or DM. Figure~\ref{fig:galpropmaps} (second, and third panels of the bottom row) shows  the predicted IC morphology for the Galactic bulge, and DM distribution, respectively.

\section{Background and foreground modeling}
\label{sec:backgroundtemplates}
In this section we describe the different templates used to model the background and foreground emission in the direction of the GC. For this we use two approaches. First, we make predictions using \textsc{galprop v56}. Second, we use phenomenological maps whose spectra are extrapolated to match \textit{Fermi}-LAT results. The misidentified CR background is obtained from detailed simulations made publicly available by the CTA consortium.

\begin{figure*}
    \begin{center}
    \includegraphics[scale = 0.50]{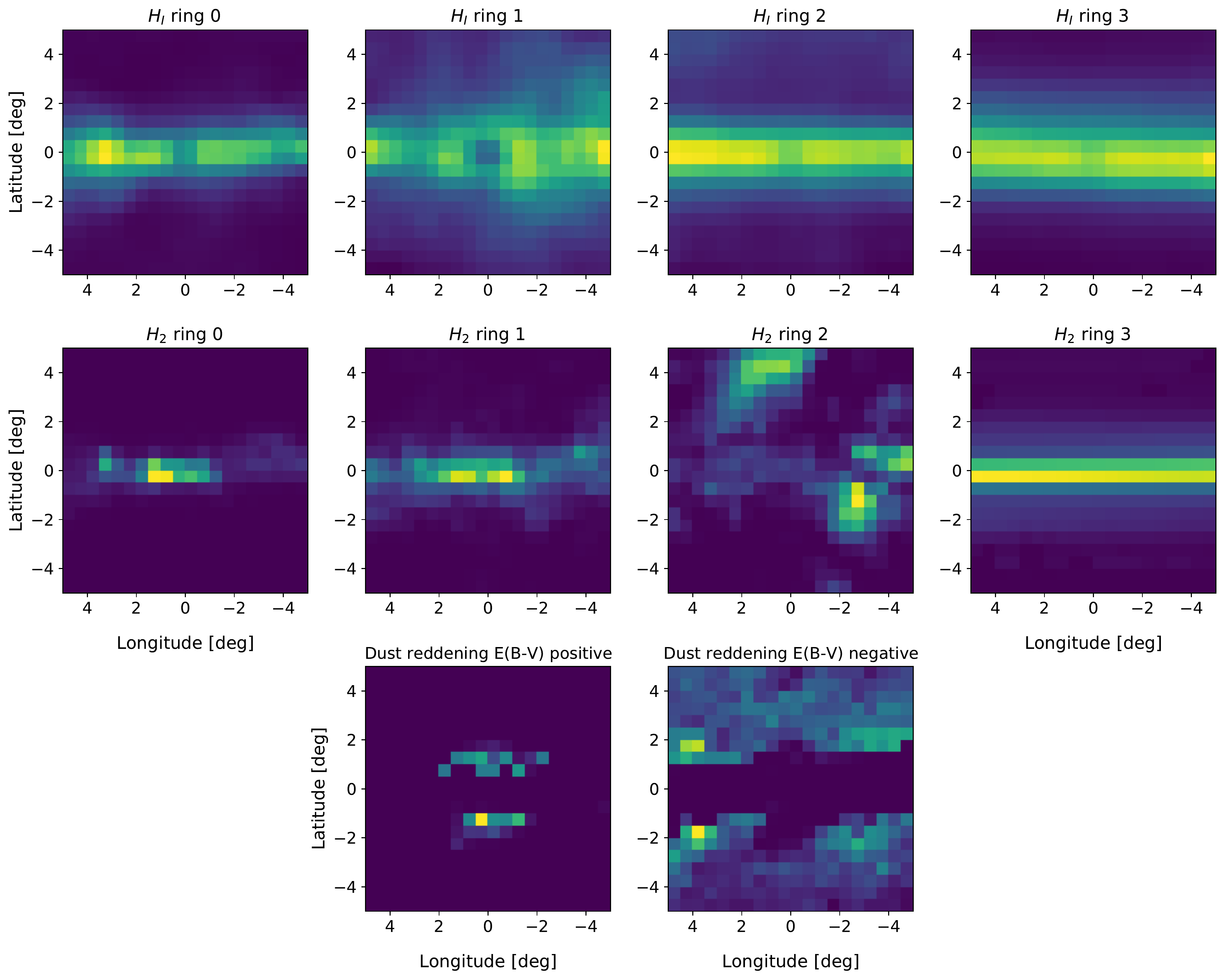}
    \caption{The morphological maps of the gas-correlated $\gamma$-ray emission constructed in~\citet{Macias:2016nev}. The maps are divided into a atomic (top panels), molecular hydrogen (middle panels), and residual dust (bottom panels) maps which trace the dark neutral material. The hydrogen maps are divided into the same four Galactocentric rings of Fig.~\ref{fig:galpropmaps}.}
    \label{fig:hydro_maps}
    \end{center}
\end{figure*}

\subsection{Irreducible isotropic $\gamma$-ray background}
The brightest source of extended $\gamma$-ray emission in the direction of the Galactic bulge corresponds to CR protons and $e^{-}$s that are misidentified as $\gamma$ rays. The interaction of highly energetic CR protons (or heavier nuclei) with the Earth's atmosphere produces showers of neutral hadrons which subsequently decay to $\gamma$ rays, an important fraction of which cannot be distinguished from astrophysical $\gamma$ rays due to the finite rejection power of CTA~\citep{Acharyya:2020sbj,Rinchiuso:2020skh}. Additionally, energetic CR $e^{-}/e^{+}$ induce electromagnetic showers that are very similar to those produced by $\gamma$ rays of astrophysical origin. 

The main feature of the resulting irreducible $\gamma$-ray background is its isotropy. Even though this component is stronger than any other in our region of interest (RoI), the goal is that by using a morphological template fit analysis, the impact of this component can be significantly reduced. We compute the template associated to this background by using dedicated simulation studies performed by the CTA consortium (see Sec.~\ref{sec:sensitivity} for details).

\subsection{Galactic diffuse emission models}
\label{subsec:GDEmodels}

The interaction of energetic CRs with interstellar gas, ambient photons, and Galactic magnetic fields produces the brightest source of $\gamma$-rays in \textit{Fermi}-LAT data. This Galactic diffuse emission (GDE) component is very difficult to model due to significant uncertainties in the interstellar gas column density maps, CR spatial/spectral distributions, and the ISRF model.   

In this study, we use two different models for the GDE. The first model, referred to as ``GDE model 1'', corresponds to one of the representative models constructed in \cite{Johannesson:2018bit}. The second model, called ``GDE model 2'', assumes the hydrodynamic gas maps constructed in \cite{Macias:2016nev}, and the IC templates used in \cite{Abazajian:2020tww}.     

Specifically, ``GDE model 1'' assumes the same propagation parameter setup shown in Table~\ref{tab:galpropsetup}, but we note that this model does not include the population of Galactic bulge MSPs that accounts for the GCE. We run \textsc{galprop v56} in its 3D mode, and divide all the predicted TeV-scale $\gamma$-ray maps in four Galactocentric rings. Since the bremsstrahlung and hadronic $\gamma$-ray maps share the same spatial morphology, we combine these two components into one (gas-correlated $\gamma$-ray emission). Dividing the IC/gas-correlated components in different rings allows us to account for the systematic uncertainties associated to the interstellar medium properties and to the somewhat uncertain spatial variation of the CRs. 
To obtain the flux normalization of each GDE component, we fitted the \textsc{galprop} maps to \textit{Fermi}-LAT observations of the inner $15^\circ \times 15^\circ$ of the GC~\citep{Macias:2016nev}, separately varying the IC and gas-correlated rings.\footnote{This is similar to the method used in \cite{Rinchiuso:2020skh}, except that here, the GDE components are divided in different rings.} Figure~\ref{fig:galpropmaps} (first and second rows) shows the predicted IC, and gas-correlated ring templates at $E=11.2$ TeV. 

For our alternative ``GDE model 2'', we use hydrodynamic maps of atomic and molecular hydrogen, in addition to dust residuals tracing dark neutral material. The main motivation of the hydrodynamic method is to reduce biases present in the standard gas maps~\citep{Ackermann:2012}. In particular, the construction of gas maps requires a model for the gas clouds' velocities in order to obtain their position with respect to the GC. 
However, the gravitational potential of the Galactic bar induces highly non-circular motion of interstellar gas in the GC region. 
To overcome this difficulty, while the standard gas maps~\citep{Ackermann:2012} generally assume pure circular orbits for the gas, for line-of-sight directions which have $\lvert l \rvert < 15^\circ $ an interpolation method must be used to obtain the gas distribution. 
Here, instead, we use the hydrodynamic method \citep{Pohl2008,Macias:2016nev} which makes direct predictions for the gas velocities in the region $\lvert l \rvert < 15^\circ $, thus providing kinematic resolution in our RoI. Figure~\ref{fig:hydro_maps} displays all the hydrodynamic templates assumed in this work, which follow the same annular subdivisions of our \textsc{galprop} templates (Fig.~\ref{fig:galpropmaps}).

Based on detailed statistical tests with \textit{Fermi}-LAT observations of the GC, \cite{Macias:2016nev,Macias:2019omb} and \cite{Buschmann:2020adf} established that the hydrodynamic maps provide a better fit to the data than the standard gas maps. Although in this article we are introducing these  maps with the purpose of estimating the systematic uncertainties in the GDE model, we expect that the hydrodynamic gas maps will be extremely useful once CTA~\citep{Acharyya:2020sbj} performs the GC survey.  

In order to have a physically motivated spectrum for ``GDE model 2'', we assume the same spectra obtained for each ring of the gas-correlated components in ``GDE model 1''. The impact of this assumption in our uncertainties estimates is expected to be small given that the gas-correlated maps have very different spatial morphologies, the templates are divided in rings, and the simulated CTA data is fitted using an analysis procedure that works bin-by-bin in energy.

\subsection{The Low Latitude Fermi Bubbles Template}
\label{subsec:FBs}

The \textit{Fermi} bubbles (FBs) are giant $\gamma$-ray lobes that were discovered in \textit{Fermi}-LAT data \citep{Su_etal:2010} using template fitting techniques. The FBs stretch out to high latitudes above and below the Galactic plane ($\lvert b \rvert \approx 55^\circ$), while their base~\citep{Herold:2019pei} is positioned slightly off-set in longitude from the GC ($l\approx-5^\circ$). 
They have an approximately uniform intensity, except for the so-called ``cocoon'' region in the southern FBs lobe~\citep{Su_etal:2012, Fermi-bubbles:2014sfa}. At latitudes $\lvert b
\rvert \geq 10^\circ$, the FBs intensity is well described by a flat power-law of the form $dN/dE\propto E^{-2}$, which  softens significantly at energies larger than 100 GeV~\citep{Fermi-bubbles:2014sfa}. At latitudes $\lvert b \rvert < 10^\circ$~\citep{Acero:2016qlg,TheFermi-LAT:2017vmf,Storm:2017arh,Herold:2019pei}, the FBs spectrum also follows a flat power-law but, peculiarly to this region, with no evidence for softening nor an energy cutoff up to energies approximately 1 TeV. 

There is still no consensus on the origin of the FBs. Several possibilities that have been discussed in the literature include: recent explosive outbursts~\citep{Su_etal:2010} from the supermassive black hole Sgr A$^\star$, and sustained nuclear processes~\citep{Crocker:2014fla} such as star formation activity from the GC. These scenarios require either IC, or hadronic gamma-ray emission to explain the observed spectra.

In this work, we model the FBs using a similar approach to the one used by~\cite{Rinchiuso:2020skh}. In particular, we assume the best-fit low-latitude FBs spectrum in \cite{TheFermi-LAT:2017vmf} and then extrapolate it to energies above 1 TeV. We consider two different sets of normalization and energy cutoff, named ``FB min'' and ``FB max'' (see Table~\ref{tab:FBmodels}). 
These are chosen such that the FBs measurements~\citep{TheFermi-LAT:2017vmf} fall within the two models, and they do not overshoot the H.E.S.S diffuse observations of the inner $\approx 0.2^\circ - 0.5^\circ$ of the Galaxy~\citep{Abramowski:2016mir}. In addition, for the spatial morphology of the FBs we use the map obtained by \citet{Macias:2019omb}. That analysis used an inpainting method to correct for artifacts introduced by the point source mask in \cite{TheFermi-LAT:2017vmf}. \cite{Macias:2019omb} validated the inpainted FBs map with a series of statistical tests. This map is shown in the left bottom corner of Fig.~\ref{fig:galpropmaps}.

\begin{table}
\begin{center}
\begin{tabular}{cccc}
\hline
\hline
Model & $N_0$ [TeV$^{-1}$cm$^{-2}$s$^{-1}$sr$^{-1}$] & $\Gamma$ & $E_{\rm cut}$ [TeV]\\
\hline
FB max & $1\times10^{-8}$ & 1.9 & 20\\
FB min & $0.5\times10^{-8}$ & 1.9 & 1\\
\hline
\hline
\end{tabular}
\end{center}
\caption{FBs models considered in this work. This component is modeled with a power-law with exponential cutoff of the form $dN/dE=N_0\; (E/{\rm 1\; TeV})^{-\Gamma}\exp(-E/{E_{cut}})$. The model parameters are the same as in~\citet{Rinchiuso:2020skh}. 
}
\label{tab:FBmodels}
\end{table}

\subsection{Point Source and Galactic plane masks}
\label{subsec:pointsources}

We model the point sources in the RoI following the same approach introduced in \cite{Rinchiuso:2020skh}. Namely, we select all the high energy point sources included in the third {\it Fermi} high energy catalog (3FHL)~\citep{TheFermi-LAT:2017pvy} that lie within our RoI ($\lvert l \rvert \leq 5^\circ$, $\lvert b \rvert \leq 5^\circ$), and for which the spectrum is given by a simple power law. We note that point sources observed to have an energy cutoff, at GeV-scale energies, in the 3FHL catalog are not included in our analysis.

Our approach is to mask the aforementioned point sources to avoid potential biases due to extrapolations. We use a disk of radius 0.25$^\circ$ centered at the best-fit position of each selected 3FHL point source.\footnote{Note that the $0.25^\circ$ masks end up being just one pixel mask due to the low resolution of the other astrophysical maps.} 
In addition, we mask the extended TeV source HESS J1745-303 -- one of the brightest sources in our RoI --
using a disk of radius 0.4$^\circ$. Since the angular resolution of CTA will be smaller than 0.1$^\circ$, we anticipate a negligible effect of potential photon leakage. 
In total, we mask five point sources in our region of interest, representing a reduction of approximately $2\%$ of our sky region.

Following ~\cite{Rinchiuso:2020skh}, we mask the Galactic plane region limited by $\lvert b \rvert \leq 0.3^\circ$, which reduces our sky region by an additional $\approx 6.5\%$. This is to avoid several bright TeV-scale point-like and extended sources in the Galactic plane. Note that even though we mask the Galactic plane, it is still important to include the nuclear bulge map in our \textsc{galprop} MSPs simulations. Energetic CR $e^{\pm}$ propagate over much greater distance scales than the size of our plane mask. Our total mask is shown in the bottom right corner of Fig.~\ref{fig:galpropmaps}.

\begin{figure*}
    \begin{center}
    \includegraphics[scale = 0.55]{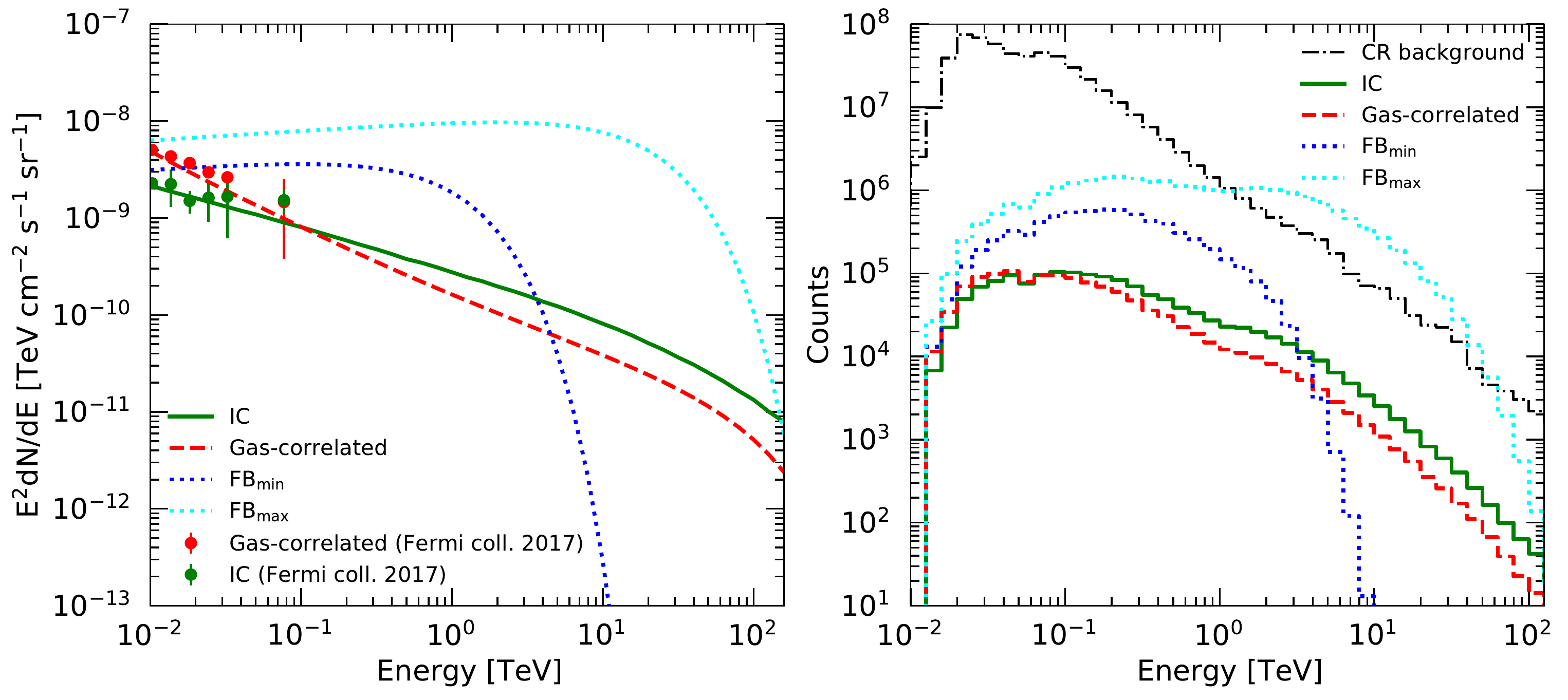}
\caption{The predicted $\gamma$-ray spectra for various Galactic diffuse emission components and their corresponding CTA count rates for 500 hours of observations of the inner $10^\circ \times 10^\circ$ of the Galactic center region. The left panel displays the bin-by-bin fluxes of the IC (green points) and gas-correlated (red points) components of the Galactic diffuse emission measured by \textit{Fermi}-LAT~\citep{TheFermi-LAT:2017vmf}. The green solid and red dashed lines are the spectra predicted by \textsc{galprop V56}. These are normalized to match the \textit{Fermi}-LAT observations at GeV-scale energies. The blue dotted line corresponds to the best fit low-latitude Fermi bubbles spectra in~\citep{TheFermi-LAT:2017vmf} extrapolated to higher energies (this is our FB$_{\rm min}$ model in Table~\ref{tab:FBmodels}). The cyan dotted line is the maximum possible emission that the FBs can take at TeV-scale energy [FB$_{\rm max}$ model in Table~\ref{tab:FBmodels}---this is the same as in Fig.1 of~\citet{Rinchiuso:2020skh}]. The right plot shows the result of convolving the Galactic diffuse emission components in the left panel with the CTA instrument response functions. The irreducible cosmic ray background (CR background) is also shown, see Sec.~\ref{subsec:expectedcounts} for details.}\label{fig:counts_spectra}
    \end{center}
\end{figure*}

\section{Sensitivity Analysis}
\label{sec:sensitivity}

We adopt the latest publicly available instrument response function (IRF) that is adequate for GC observations (\textit{CTA-Performance-prod3bv1-South-20deg-average-50h.root}\footnote{\url{http://cta.irap.omp.eu/ctools/users/user_manual/response.html}}). This contains information of the energy-dependent effective area, point spread function, energy resolution, and irreducible $\gamma$-ray background (in the energy range 10 GeV--100 TeV). The IRF utilized here was constructed by the CTA team~\citep{CTA_IRFs:2015} using a suite of dedicated Monte Carlo simulations. These assume an array of detectors composed of 4 large-size telescopes (23 m diameter and sensitive to photons in the 20--150 GeV range),  24 medium-size telescopes (11.5 m diameter and sensitive to the 150 GeV--5 TeV range) and 70 small-size telescopes (4 m diameter and sensitive to the highest energies). At TeV energies, CTA has an energy resolution as good as approximately 5\%.

For our analysis we assume the most favorable observation conditions; a 500 h on-axis observation from the southern site\footnote{This could be obtained with an optimized observation strategy planned by the CTA consortium~\citep{Acharyya:2020sbj}.} at a mean zenith angle of $20^\circ$. Additionally, we consider events with energies between 16 GeV and 158 TeV. The RoI is selected to be a square of size $10^\circ \times 10^\circ$---centered at Galactic coordinates $(l,b)=(0^\circ, 0^\circ)$---which is further binned into pixels of size $0.5^\circ \times 0.5^\circ$. This is the same binning scheme introduced in \cite{Rinchiuso:2020skh}. These authors showed that the choice of bin size had no impact on the results given the high photon statistics obtained in each spatial bin. We note that all our background and signal models constructed with \textsc{galprop v56} also have a resolution of $0.5^\circ \times 0.5^\circ$. Increasing the resolution further is limited by computational costs.  

\subsection{Computation of the Expected Photon Counts }
\label{subsec:expectedcounts}

In order to get the expected counts for a given sky model, we convolve the signal/background templates with the CTA IRFs described above. The convolution is done with the \textsc{Gammapy}~\citep{gammapy:2017,gammapy:2019} analysis tools,\footnote{\url{https://gammapy.org/}} and the model templates are shown in Fig.~\ref{fig:galpropmaps} and \ref{fig:hydro_maps}.

Since the CTA's point spread function is smaller than our pixel size ($0.5^{\circ} $ x $ 0.5^{\circ}$), it can be neglected in our calculations. The function that describes the expected counts $\Phi^m_{ij}$ for a certain astrophysical component $m$, at the $i^{th}$ longitudinal, $j^{th}$ latitudinal, and $\Delta E$ energy bin can then be written as

\begin{equation}\label{eq:expectedcounts}
    \Phi^m_{ij} = T_{obs} \int_{\Delta E} dE_{\gamma} \int_{0}^{\infty} dE^{'}_{\gamma}  \frac{d\phi^{m}_{ij}}{dE_{\gamma}} A^{\gamma}_{eff}(E^{'}_{\gamma}) D(E_{\gamma},E_{\gamma}^{'}),
\end{equation} 
where $A^{\gamma}_{eff}(E_\gamma)$ is the energy-dependent effective area, $D(E_{\gamma},E_{\gamma}^{'})$ is the energy dispersion function, and $d\phi^{m}_{ij}/dE_{\gamma}$ is the incoming flux spectrum from the $m$ source.\footnote{The irreducible CR background is formally included as an additional component $m$ in Eq.~\ref{eq:expectedcounts}.} The whole function is integrated over the energy bin $\Delta E_{\gamma}$, and multiplied by the total observation time $T_{obs}$. The two different energies stand for the true incoming energy $E_{\gamma}^{'}$, and the reconstructed energy $E_{\gamma}$. Assuming a homogeneous sky exposure in our RoI, we set $T_{obs}=500$ hours in our analysis. 

As an example of the IRFs convolution results, we show in Fig.~\ref{fig:counts_spectra} (left) the predicted spectra for each of the background model components, and in Fig.~\ref{fig:counts_spectra} (right) the same background templates after convolution with the CTA IRFs. We have carefully checked that our pipeline reproduces well previous works in the literature [e.g.,~\citep{Rinchiuso:2020skh}].

\subsection{Template fitting procedure}
\label{sub:templatefitting}

The conventional analysis method for TeV-scale $\gamma$-ray observations involves selecting two regions of the sky with approximately the same backgrounds, 
but different expected signals. The region with larger signal is called the ``ON'' region, while the other the ``OFF'' region. In this approach, the hypothesis testing is done using a test statistic (TS) defined as the difference in photon counts between the ON and OFF regions. However, the study by \cite{Silverwood:2014yza} demonstrated that the template-fitting procedure---which is the standard analysis method for e.g., \textit{Fermi}-LAT data---greatly improves the CTA sensitivity to extended GC signals because it allows for the full exploitation of the morphological differences between background and signal templates. 

In this work, we adopt a template-fitting approach to study CTA sensitivities to a putative IC signal from Galactic bulge MSPs. In particular, we divide the mock data in 11 bins logarithmically spaced from $16$ GeV to $158$ TeV. This makes the bins larger than the energy resolution of the CTA data. 
For each independent energy bin $\Delta E$, we define the likelihood function as   

\begin{equation}\label{eq:likelihoodfunc}
   \mathcal{L}(\boldsymbol{\mu}\lvert \boldsymbol{n}) = \prod_{ij}\frac{\mu_{ij}^{n_{ij}} e^{-\mu_{ij}}}{n_{ij}!}
\end{equation}
where $\boldsymbol{n}=\{n_{ij}\}$ is the simulated data, and the model $\boldsymbol{\mu}=\{\mu_{ij}\}$ is a linear combination of model templates

\begin{equation}\label{eq:superposition}
    \mu_{ij}(\boldsymbol{\alpha}) = \sum_{m} \alpha_{m} \Phi^m_{ij}, 
\end{equation}
with $\Phi^m_{ij}$ as given in Eq.~\ref{eq:expectedcounts}, and the flux normalizations $\alpha_m$ are the fitting parameters. The optimization is done with the \textsc{minuit}~\citep{Nelder:1965zz} algorithm contained in the \texttt{iminuit} package.\footnote{\url{https://iminuit.readthedocs.io/en/stable/}} 

For each independent energy bin, we simultaneously fit the flux normalizations of all the background and signal components within our $10^\circ \times 10^\circ$ and $|b|\geq 0.3^\circ$ RoI. The normalizations of the sources considered are insensitive to the spectral shape assumed at each energy bin due to the bins being small.  This fitting approach, also known as a bin-by-bin analysis, has been widely used in the analysis of \textit{Fermi}-LAT data~\cite[e.g.,][]{Ackermann:2015zua}. The advantage of this method over the more traditional broad-band analysis~\citep[e.g.,][]{Rinchiuso:2020skh,Acharyya:2020sbj}, is that the bin-by-bin method is much less affected by assumptions about the spectral shape of the model templates. 

Since CTA is not in operation yet, we create synthetic data for each energy bin and each component. We do this by drawing from a Poisson distribution with mean $\mu_{ij}(\alpha_m)$, and then summing the resulting maps to obtain the total mock data set for an energy bin.

We evaluate the significance of the MSPs hypothesis at each energy bin $\Delta E_k$ using a test statistic defined as

\begin{equation}\label{eq:TSbin}
    {\rm TS_k} = -2\ln{\left(\frac{\mathcal{L}(\boldsymbol{\mu_0, \hat{\theta}}\lvert \boldsymbol{n})}{\mathcal{L}(\boldsymbol{\hat{\mu}, \hat{\theta}}\lvert \boldsymbol{n})}\right)}
\end{equation}
where $\boldsymbol{\mu_0}$ are the normalizations of the background-only hypothesis, and $\boldsymbol{\hat{\mu}}$ and $\boldsymbol{\hat{\theta}}$ are the best-fit parameters under the background plus MSPs hypothesis. The total TS of the MSPs IC template can be obtained as $\mbox{TS}=\sum_{k=1}^n \mbox{TS}_k$ where $\mbox{TS}_k$ is given in Eq.~\ref{eq:TSbin}, and the sum runs up to $n=11$ (number of energy bins). In our case, the MSPs IC template has 11 degrees of freedom (flux norm at each energy bin). Hence, in order to get the p-value of this template we are required to use the mixture distribution formula~\citep{Macias:2016nev}

\begin{equation}\label{eq:pvalue}
p({\rm TS})= 2^{-n}\left(\delta({\rm TS}) +\sum_{k=1}^{n} \binom{n}{k} \chi_{k}^2({\rm TS})\right)
\end{equation}
where $\binom{n}{k}$ is the binomial coefficient with $n=11$, $\delta$ is the Dirac delta function, and $\chi_{k}^2$ is a $\chi^2$ distribution with $k$ degrees of freedom. We can compute the detection significance in $\sigma$ units  
corresponding to the addition of one new MSP IC norm parameter
by using the total TS value and the p-value shown in the above equation. Specifically, we do this with the following recipe~\citep{Macias:2016nev}
\begin{equation}\label{eq:numberofsigmas}
\mbox{Number of $\sigma$}\equiv \sqrt{\rm InverseCDF\left(\chi_1^2,{\rm CDF}\left[p(\mbox{TS}),\hat{{\rm TS}}\right]\right)}
\end{equation}
where (InverseCDF) CDF is the (inverse) cumulative distribution. The first argument of each of these functions is the distribution function and the second is the value (inverseCDF) at which the CDF is evaluated. 
The total TS value is denoted by $\hat{\rm TS}$. From Eq.~\ref{eq:numberofsigmas} we obtain that a 5$\sigma$ detection corresponds to $\mbox{TS}=41.1$. It follows that we may  
claim a detection of the IC template for total TS values larger than this threshold value.

\section{Results}
\label{sec:results}

Having introduced the fitting procedure and explained how we create the CTA simulated data, we now present the results of our CTA sensitivity analysis.
Specifically, we start by working out the number of independent gas-correlated rings for which we can get stable fits for the background only hypothesis. Then, we present the minimum IC (and $e^\pm$) luminosities necessary for a reliable MSPs signal detection, and the ability of the method to accurately distinguish between the MSPs and dark matter origin of the radiating $e^\pm$. Finally, we show the impact of the Galactic diffuse emission model uncertainties on our results.

\subsection{Validation of the Galactic diffuse emission model on simulated CTA data}
\label{subsec:validationGDE}

To the best of our knowledge, this is the first time that a Galactic diffuse background model divided in multiple galactocentric rings (see Fig.~\ref{fig:galpropmaps}) is used for TeV-scale $\gamma$-ray analyses. Though this method has been implemented with great success in studies of \textit{Fermi} data [e.g.,~\citet{Fermi-LAT:4FGL}], it is not {\it a priori} obvious that the same technique can be applied to CTA, given its smaller field of view.

To address this concern
we performed a fitting procedure where we carefully checked for potential degeneracies between the different gas-correlated and IC rings shown in Fig.~\ref{fig:galpropmaps}. From this test, we obtained that the morphological differences between the four galactrocentric IC emission maps were not significant enough for the pipeline to distinguish them in the simulated data. This is because most of these spatial differences lay in the parts that are further away from our RoI (central $10^\circ\times10^\circ$ and $|b|\geq0.3^\circ$ of the Galaxy). We therefore decided to
combine the four IC rings into one single map and only keep the gas-correlated emission
maps split into four galactocentric rings.

It is worth noting that using different IC rings for GC analyses could still be viable with a different observational strategy. In particular, the GC survey plan proposed in Fig.1 of~\citet{Acharyya:2020sbj} covers a region that is almost two times larger than the RoI assumed in our work. Another interesting possibility could be to use the CTA divergent pointing mode~\citep{Gerard:2015gge} in which CTA can survey a region as large as $20^\circ \times 20^\circ$ [see for example the ``Deep exposure scenario'' proposed in~\cite{Coronado-Blazquez:2021avx}]. We leave these interesting alternatives for future studies, and stick only with our survey strategy, which was suggested in~\citet{Rinchiuso:2020skh}.

\subsection{Sensitivity to the IC signal produced by an unresolved population of MSPs in the GC}
\label{subsec:sensitivityICMSPs}

Given that the systematic uncertainties in the GDE model is one of the most difficult problems for GeV-scale $\gamma$-ray analyses of the GC, it is reasonable to assume that this component will also be very challenging for forthcoming CTA observations of similar sky regions. With this in mind, we started our sensitivity analysis by testing how well our pipeline recovers the properties of the simulated MSP IC signal in different case scenarios. In particular, we considered various alternative GDE models, different signal spectra and spatial morphologies, and we further employed a method to study the impact of mismodeling the Galactic diffuse backgrounds. Details of each of our GDE model components are given in Sec.~\ref{sec:backgroundtemplates}. As for the MSPs IC signal, we considered a range of physically reasonable injection spectra presented in Table~\ref{tab:mspspectrum}.

\begin{figure*}
    \begin{center}
    \includegraphics[scale = 0.41]{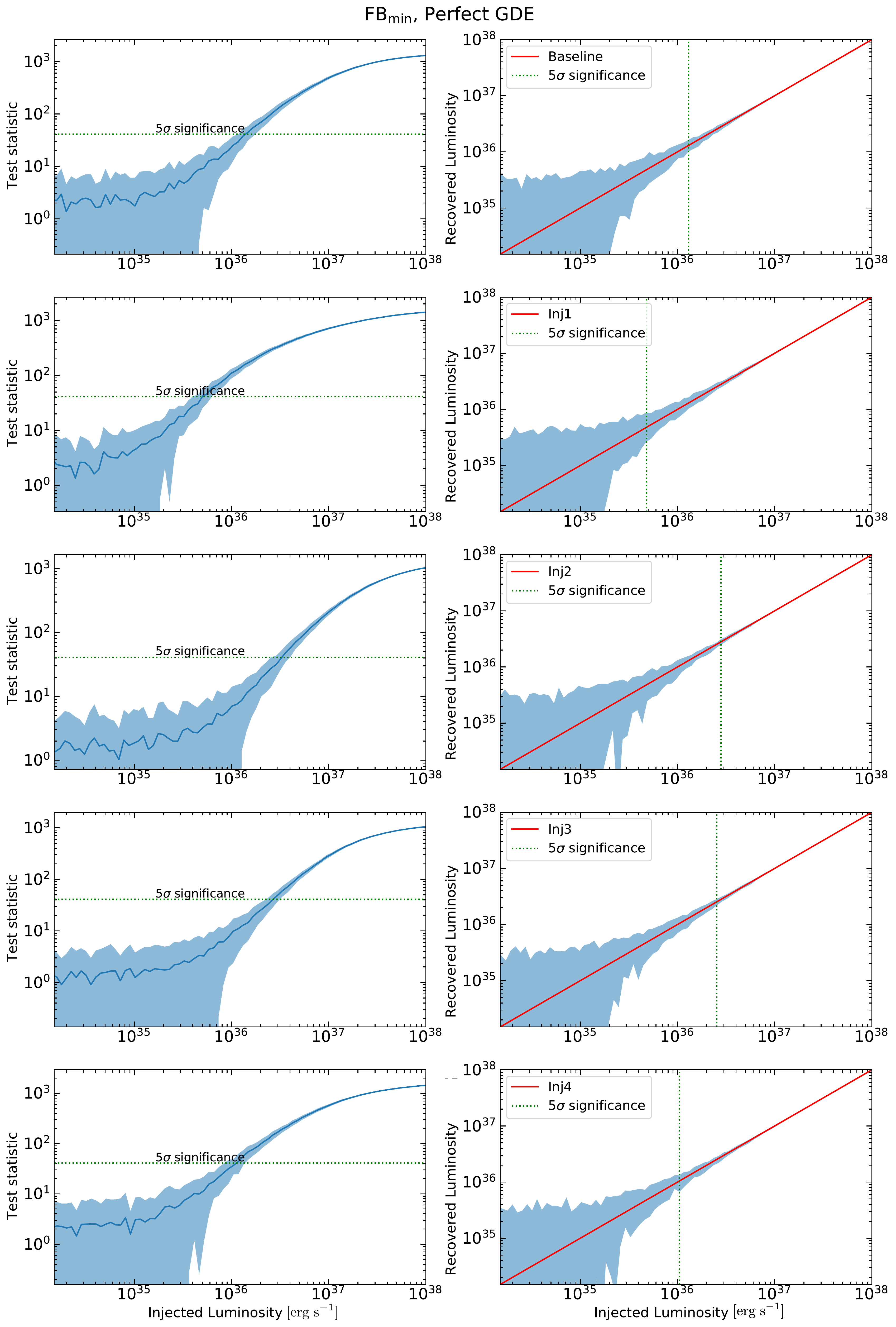}
    \caption{The CTA sensitivity to the IC signal from an unresolved population of MSPs tracing the distribution of stellar matter in the Galactic bulge. These tests assume a perfect knowledge of the Galactic diffuse emission. The latter is modeled with a combination of the FB$_{\rm min}$ model, and ``GDE model 1'' (see also Sec.~\ref{subsec:GDEmodels})  
    . \textbf{Left panels:} Detection significance (Test statistic) of the MSPs' IC signal for a given IC injection luminosity. Each row assumes a different $e^\pm$ spectrum model (Inj1,..., Inj4) shown in Table~\ref{tab:mspspectrum}. The $5\sigma$ detection threshold is displayed as a green dotted line (see Eqs.~\ref{eq:pvalue} and~\ref{eq:numberofsigmas} for details). A summary of the minimum $L_{\gamma,IC}$ required for a CTA detection is shown in Table~\ref{tab:minimumluminosities}.  The blue solid line represents the mean of the results, while the light blue region gives its variance. \textbf{Right panels:} Comparison of the recovered signal luminosity with the injected one. The diagonal red line represents the ideal case in which the extracted signal matches the injected signal perfectly. The blue region shows the recovered IC signals. We used  5000 realizations of synthetic data that contains irreducible cosmic-ray background photons plus astrophysical $\gamma$ rays sampled from the aforementioned diffuse model. 
    }\label{fig:Injection_perfectGDE}
    \end{center}
\end{figure*}

\begin{figure*}
    \begin{center}
    \includegraphics[scale = 0.48]{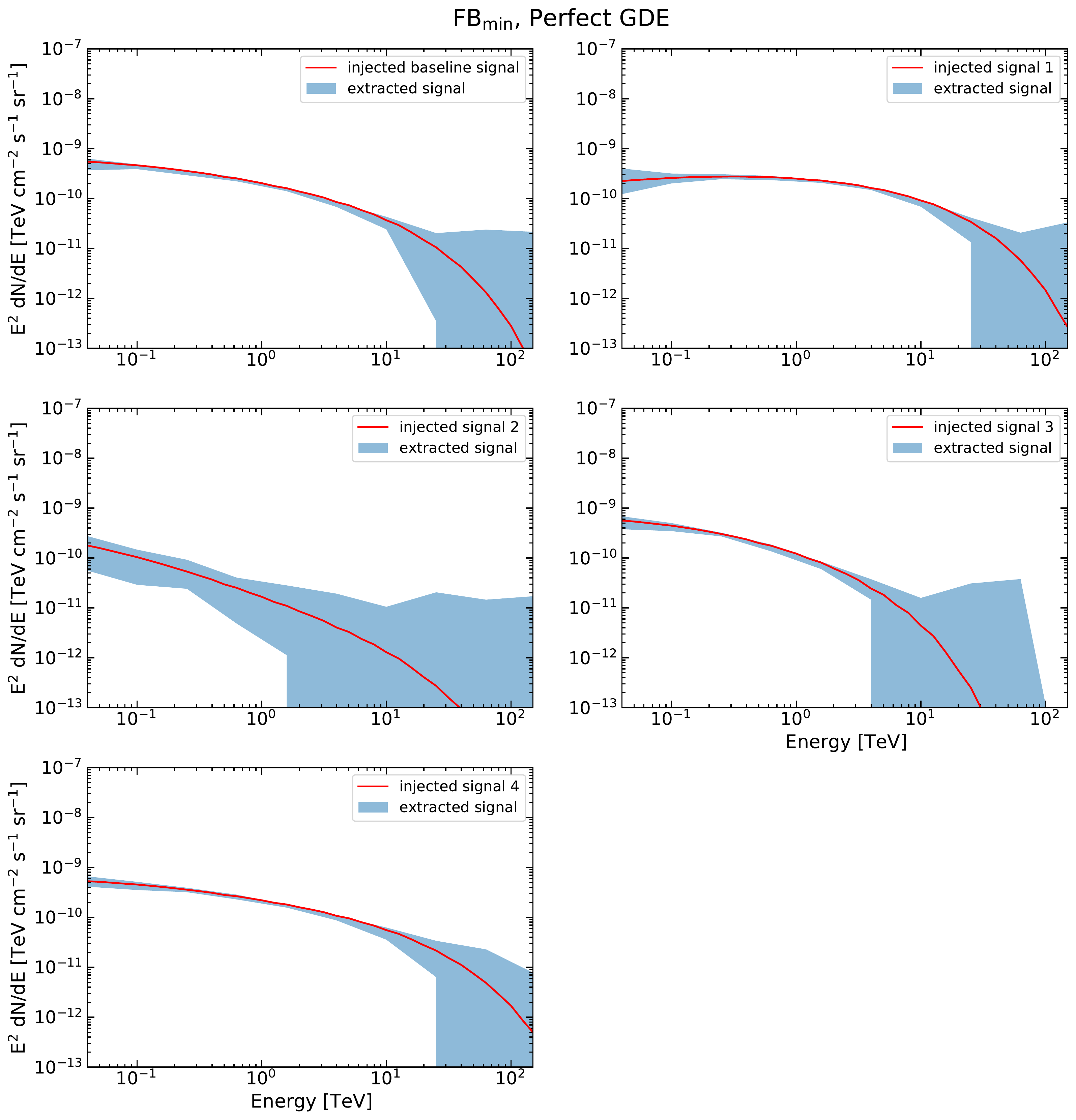}
    \caption{Minimum flux to detect the IC signal from a putative population of MSPs responsible for the GCE. The red line shows the injected IC signal while the blue regions display the 68\% confidence intervals on the recovered signal. The normalization of the spectra corresponds to the $\gamma$-ray luminosity that would be detected with $5\sigma$ significance (see also the green dotted line in Fig.~\ref{fig:Injection_perfectGDE}) for 500 h of CTA observations of the central $10^\circ\times10^\circ$ and $|b|\geq0.3^\circ$ of the Galaxy. Higher fluxes than the ones displayed here would be detected by CTA with a statistical significance larger than $5\sigma$.      }\label{fig:RecoveredSpectrum_perfectGDEFBmin}
    \end{center}
\end{figure*}

\begin{figure*}
    \begin{center}
    \includegraphics[scale = 0.48]{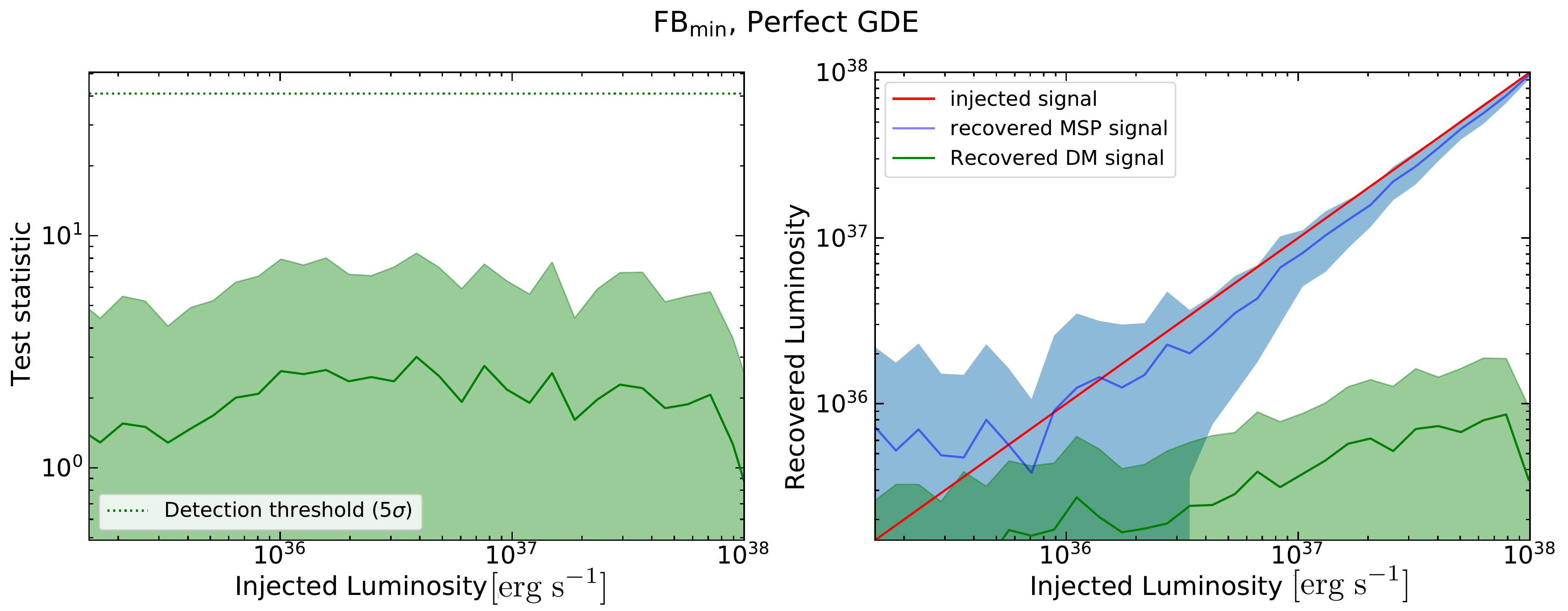}
\caption{Tests of spectro-morphological degenerecies between the MSPs IC and DM IC templates with simulated CTA observations of the GC region. The astrophysical background is sampled from "GDE model 1", FB$_{\rm min}$, and the irreducible CR background (see also caption of Fig.~\ref{fig:Injection_perfectGDE}). A MSPs IC signal is injected into the mock data and subsequently a bin-by-bin fitting procedure is applied using the same background templates used in the generation of the mock data, in addition to the MSPs IC and a DM IC templates. The left panel shows the mean TS distribution (green solid line) and corresponding 68\% confidence region (gree filled area) for the DM IC template. The right panel displays the fraction of MSPs IC (blue filled area) and DM IC (green filled area) luminosities that are obtained after injection of only the MSPs IC signals of various luminosities. }\label{fig:MSPsvsDM_perfectGDE}
    \end{center}
\end{figure*}

Figure~\ref{fig:Injection_perfectGDE} shows the results of our signal recovery tests in the case where the diffuse model is given by ``GDE model 1'' plus FB$_{\rm min}$, and we assume that the backgrounds are perfectly modeled; this is accomplished by fitting the mock data with the same templates used in the generation of the simulations. This figure is made from 5000 realizations of synthetic data that contains irreducible cosmic ray background photons plus astrophysical $\gamma$ rays sampled from the aforementioned diffuse model. Each row corresponds to the five different injection models introduced in Table~\ref{tab:mspspectrum}. The luminosity of the signal that is injected (red solid line) is displayed along with the 68\% containment on the recovered luminosity (blue region) in the right hand side panels. We also include the $5\sigma$ (${\rm TS}=41.1$ for 11 degrees of freedom) detection threshold (green dotted line), representing the luminosity above which CTA would reliably detect the IC signal from the putative MSP population in the GC. As can be seen, the injected signal is successfully recovered for luminosities above that in the detection threshold. Below this threshold, the uncertainty on the recovered $\gamma$-ray luminosity increases very similarly for all the injection models that were included (see Table~\ref{tab:mspspectrum}).     

In the left panels of Fig.~\ref{fig:Injection_perfectGDE} we present the TS distributions of the signal templates. We stress that while fitting the mock data we allow all components (background and signal templates) to vary in the fits. The filled regions denote the TS distributions of the MSPs' IC signal (light blue) and their mean value (blue solid line) is also displayed. These demonstrate that with 500 h of GC (central $10^\circ\times10^\circ$ and $|b|\geq0.3^\circ$ of the Galaxy)  observations 
with the CTA
and an accurate knowledge of the astrophysical backgrounds (corresponding to the most optimistic scenario considered in this work) the technique presented here should be able to detect the IC signal from the putative GC MSP population, thereby allowing us to constrain the source populations generating the high-energy $\gamma$ rays in this sky region.

For this same case scenario, we present details of the recovered spectra in Fig.~\ref{fig:RecoveredSpectrum_perfectGDEFBmin}. Each panel in this figure shows the characteristics of the spectra that are recovered with a $5\sigma$ detection significance. The five different panels correspond to each $e^\pm$ injection model presented in Table~\ref{tab:mspspectrum}. As explained in Sec.~\ref{sec:ICfromMSPs}, we propagate such $e^\pm$ within the \textsc{galprop} V56 framework, then produce IC spectral templates (red solid lines), and lastly inject these signal maps into the simulated data. We obtained the results in this figure by fitting the mock data with a bin-by-bin fitting procedure that allows us to work out the fluxes and corresponding 68\% confidence intervals (blue filled regions) at each independent energy bin. Fluxes larger than those shown in these panels should be successfully detected by CTA, provided the parent $e^\pm$'s are injected by an unresolved population of MSPs (tracing the distribution of stellar mass in the Galactic bulge) and the Galactic diffuse background is perfectly modeled. We also present a summary of the minimum $L_{\gamma,{\rm IC}}$ (and $L_{e^\pm}$) 
required for a CTA detection in Table~\ref{tab:minimumluminosities}.

\subsection{Tests for degeneracies between the IC maps from MSPs and DM in the GC}
\label{subsec:degeneracyDMvsMSPs}

In the previous section (Sec.~\ref{subsec:sensitivityICMSPs}) we created mock data by sampling from the  ``GDE model 1'', FB$_{\rm min}$, and the irreducible CR background. We then injected various different MSP IC signals into the data and applied a fitting procedure to recover the injected signals. In particular, the fit included all the templates used in the generation of the mock data. 

In this subsection, we applied the same pipeline, except that this time we also added to the fit an IC template generated by DM emission. Namely, we injected a MSP IC signal into the mock data, and subsequently attempted to recover it by including both the MSP IC and DM IC templates in the fit. The main objective of this test is to figure out the conditions under which CTA would be able to disentangle a new extended $\gamma$-ray source in the GC based on the morphological characteristics of the IC radiation emitted by the source.

We present the results of the tests for degeneracy between these two competing hypotheses in Fig.~\ref{fig:MSPsvsDM_perfectGDE}. The left panel shows the TS distribution of the DM IC template as a function of the injected MSPs IC luminosity. In this panel, we display the mean TS values (green solid line) and their respective 68\% containment band (green filled region). As can be seen, for all the evaluated luminosities, the DM IC template was found to have ${\rm TS}\lesssim 10$ (or a statistical significance of $\lesssim 1.6\sigma$ for 11 degrees of freedom). The right panel shows the injected MSP IC luminosity versus the luminosities recovered for the MSPs IC (blue filled region) and the DM IC (green filled region) templates, respectively. It is clear that for IC luminosities larger than the minimum $L_{\gamma}$'s given in Table~\ref{tab:minimumluminosities} (see the row corresponding to FB$_{\rm min}$, mismodeling of the GDE), only a small fraction of the injected MSP IC luminosity is absorbed by the DM IC template. 

\begin{table}
  \begin{center}
    \begin{tabular}{c|c|c|c|c}
    \hline\hline
    \rowcolor[gray]{.9}  
     \multicolumn{5}{c}{ Minimum $L_{\gamma, {\rm IC}}$  for detection [erg s$^{-1}$]}   \\\hline
     \textbf{Baseline} & \textbf{Inj1} & \textbf{Inj2} & \textbf{Inj3} & \textbf{Inj4}\\\hline
      \rowcolor[gray]{.9}
     \multicolumn{5}{c}{  FB$_{\rm min}$, perfect GDE.}   \\\hline
     $1.3\times 10^{36}$ & $5.4\times 10^{35}$ & $3.3\times 10^{36}$ & $2.7\times 10^{36}$ & $1.0\times 10^{36}$\\\hline
      \rowcolor[gray]{.9}
     \multicolumn{5}{c}{ FB$_{\rm min}$, mismodeling of the GDE.}\\\hline
     $1.8\times 10^{36}$ & $7.2\times 10^{35}$ & $3.4\times 10^{36}$ & $2.8\times 10^{36}$ & $1.3\times 10^{36}$\\\hline
      \rowcolor[gray]{.9}
     \multicolumn{5}{c}{FB$_{\rm max}$, perfect GDE.}\\\hline
     $7.1\times 10^{36}$ & $7.5\times 10^{36}$ & $5.4\times 10^{36}$ & $6.1\times 10^{36}$ & $7.2\times 10^{36}$\\\hline
      \rowcolor[gray]{.9}
     \multicolumn{5}{c}{ 
     FB$_{\rm max}$, mismodeling of the GDE.}\\\hline
     $9.0\times 10^{36}$ & $9.4\times 10^{36}$ & $6.8\times 10^{36}$ & $7.8\times 10^{36}$ & $9.1\times 10^{36}$\\
     \hline\hline
    \end{tabular}
    \caption{Minimum  IC luminosity required for a $5\sigma$ significance detection with CTA. The luminosities are computed in the energy range 16 GeV and 158 TeV. The columns correspond to different MSPs $e^\pm$ spectra considered in Table~\ref{tab:mspspectrum}. The corresponding minimum $L_{e^\pm}$ can be obtained by realizing that the $L_{e^\pm}/L_{\gamma, {\rm IC}}$ ratios are: 21.0 for the \textit{baseline} injection model, 14.3 for \textit{inj1}, 124.7 for \textit{inj2}, 23.4 for \textit{inj3}, and 21.1 for  \textit{inj4}. See also Sec.~\ref{sec:efficiency} for details. For the computation of the efficiencies we assumed energies greater than 700 MeV.}\label{tab:minimumluminosities}
  \end{center}
\end{table}

\subsection{Mismodeling of the Galactic diffuse emission}
\label{subsec:GDEmismodeling}

Due to uncertainties in the Galactic diffuse background, it will be challenging for future analyses of actual CTA data to model this component perfectly~\citep{Acharyya:2020sbj}. 
We re-create this  
real-world situation
by constructing simulated data with ``GDE model 1'' and analyzing it with ``GDE model 2'', which allows us to test whether mismodeling of the GDE could originate in a false positive detection of an MSP IC signal. This methodology is inspired by a recent study on simulated {\it Fermi} data by~\citep{Chang:2019ars}, and a similar one performed by the CTA consortium~\citep{Acharyya:2020sbj} in the context of DM searches in the GC. 

\begin{figure*}
    \begin{center}
    \includegraphics[scale = 0.42]{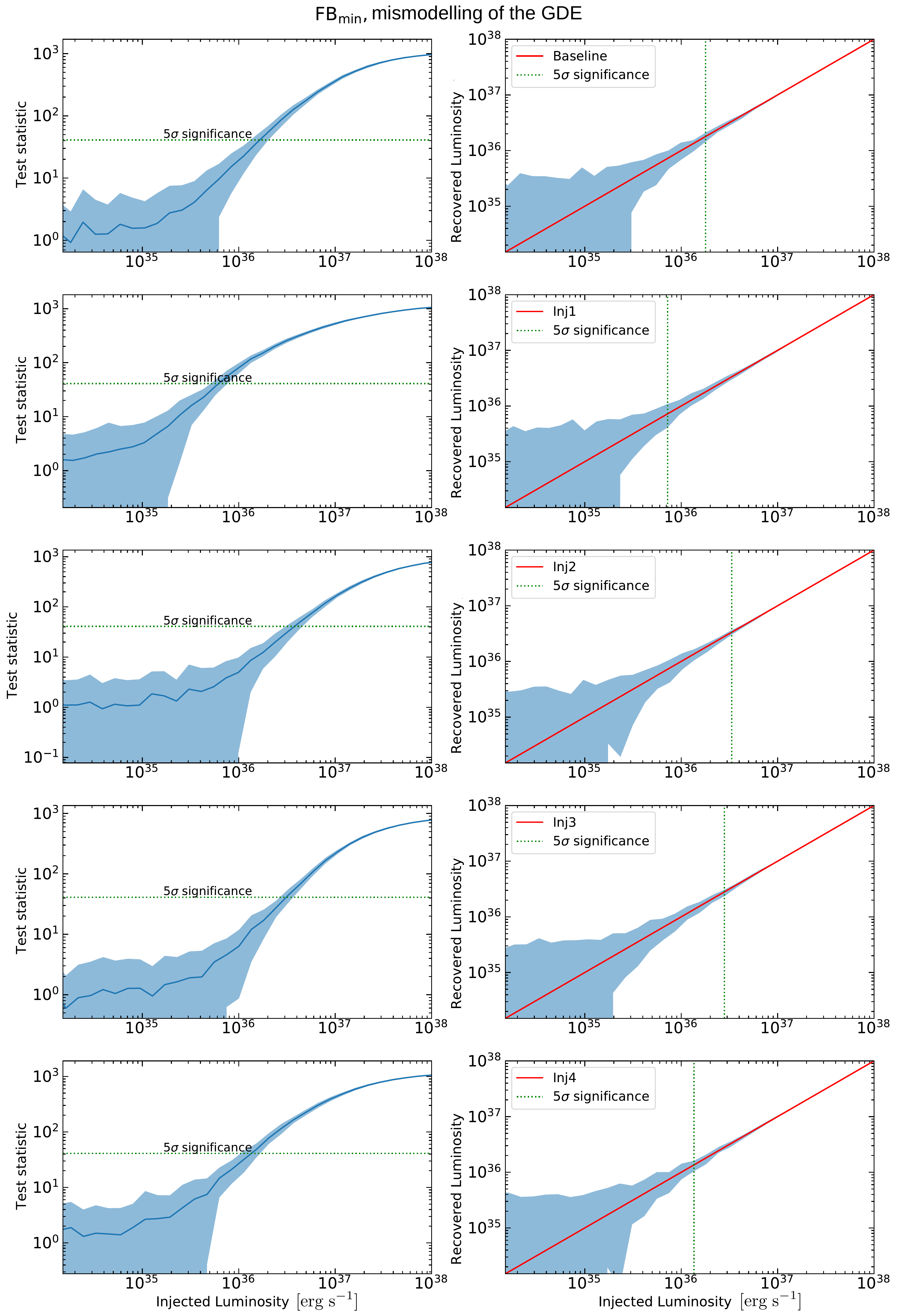}
    \caption{Same as Fig.~\ref{fig:Injection_perfectGDE}, except that here we assume mismodeling of the Galactic diffuse emission model and the FB$_{\rm min}$ model. Note that mismodeling of the GDE is mocked up by generating the data with ``GDE model 1'' and then fitting the data with ``GDE model 2''. }\label{fig:InjectionmismodelingFBmin}
    \end{center}
    \end{figure*}

\begin{figure*}
    \begin{center}
    \includegraphics[scale = 0.48]{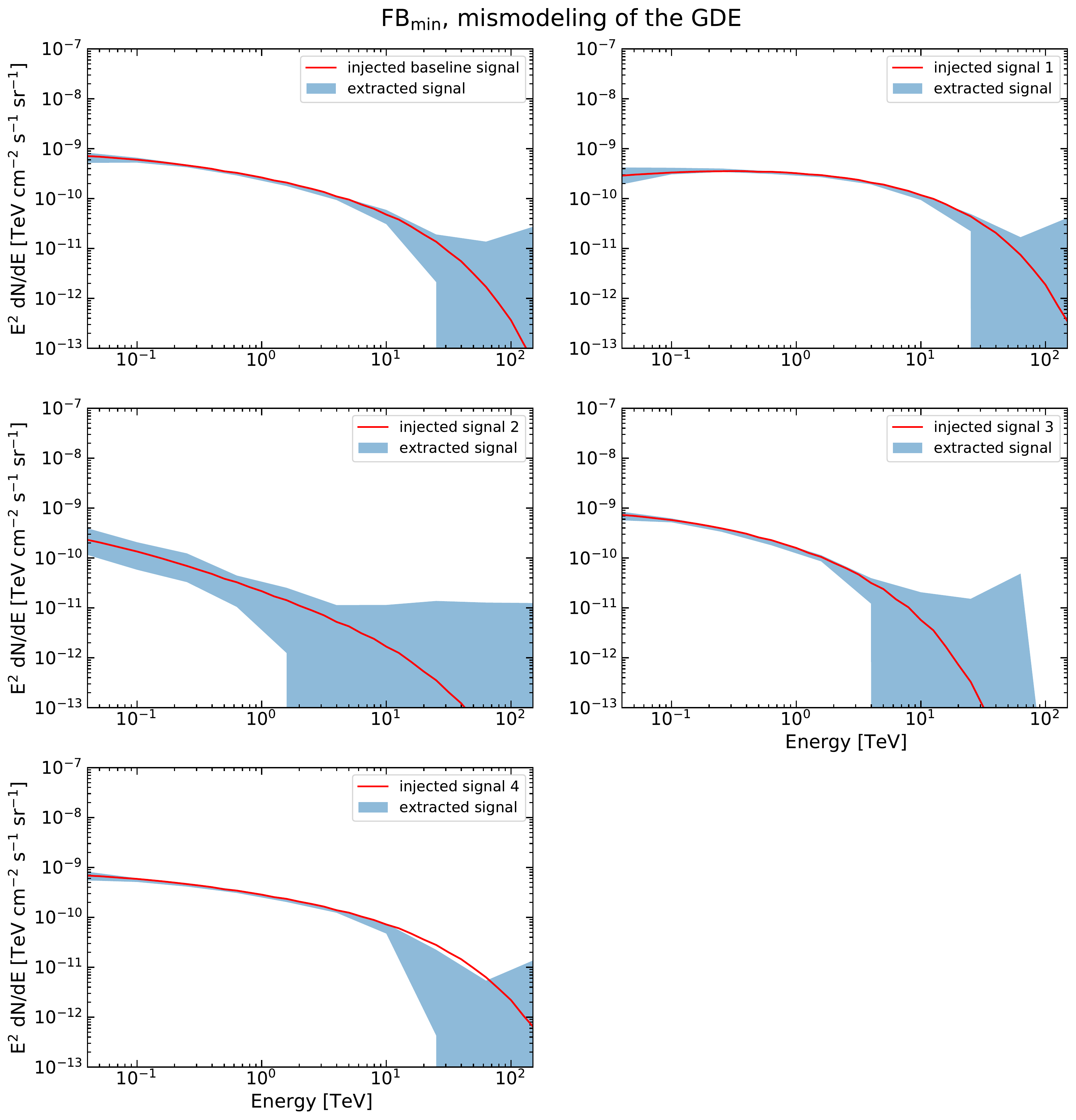}
\caption{Same as Fig.~\ref{fig:RecoveredSpectrum_perfectGDEFBmin}, except that here we assume mismodeling of the Galactic diffuse emission model. See also the green dotted line in Fig.~\ref{fig:InjectionmismodelingFBmin} for the necessary IC luminosity for a $5\sigma$ significance detection of the signal. }\label{fig:RecoveredSpectrummismodelingGDEFBmin}
    \end{center}
    \end{figure*}

\begin{figure*}
    \begin{center}
    \includegraphics[scale = 0.48]{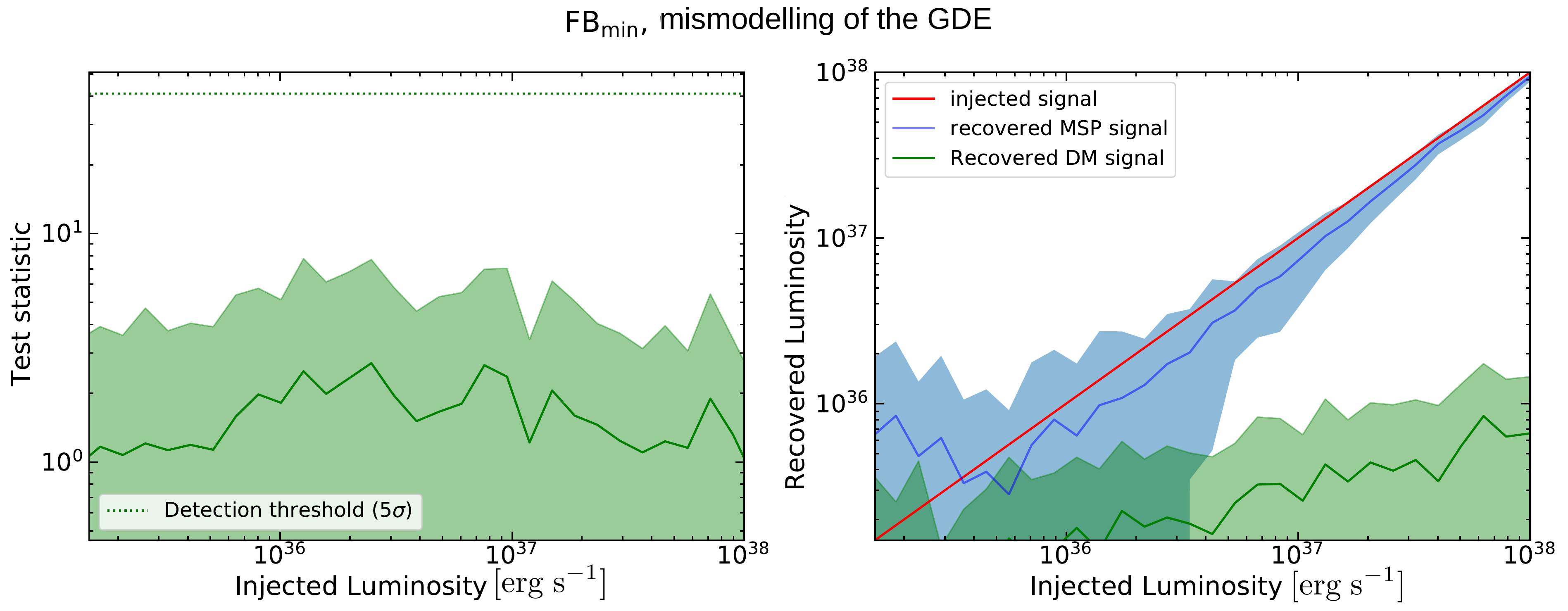}
    \caption{Same as Fig.~\ref{fig:MSPsvsDM_perfectGDE}, except that here we assume mismodeling of the Galactic diffuse emission.  }\label{fig:MSPsvsDMmismodelingGDEFBmin}
    \end{center}
\end{figure*}

We thus repeated the same analyses performed in Sec.~\ref{subsec:sensitivityICMSPs} and Sec.~\ref{subsec:degeneracyDMvsMSPs}, but this time, mimicking the mismodeling of the Galactic diffuse emission as described above. We show the results, respectively, for the signal recovery test in Fig.~\ref{fig:InjectionmismodelingFBmin}, the minimum flux required for a $5\sigma$ significance detection of the MSPs IC signature in Fig.~\ref{fig:RecoveredSpectrummismodelingGDEFBmin}, and lastly, the degeneracy tests between the MSPs and DM hypotheses in Fig.~\ref{fig:MSPsvsDMmismodelingGDEFBmin}. 

We found that, in the case where the Galactic diffuse background is mismodeled (``FB$_{\rm min}$, mismodeling of the GDE'' scenario), the CTA sensitivity to the MSPs IC signal is reduced   by approximately 38\%, 33\%, 3\%, 4\%, and 30\%, respectively, for the $e^\pm$ injections scenarios  \textit{Baseline}, \textit{Inj1}, \textit{Inj2}, \textit{Inj3}, and \textit{Inj4} (see Table~\ref{tab:mspspectrum}). A summary of the minimum $L_{\gamma, {\rm IC}}$ required for a CTA detection is given in Table~\ref{tab:minimumluminosities}.

\subsection{Impact of the low-latitude \textit{Fermi} bubbles model}
\label{subsec:FBsImpact}

\begin{figure*}
    \begin{center}
    \includegraphics[scale = 0.42]{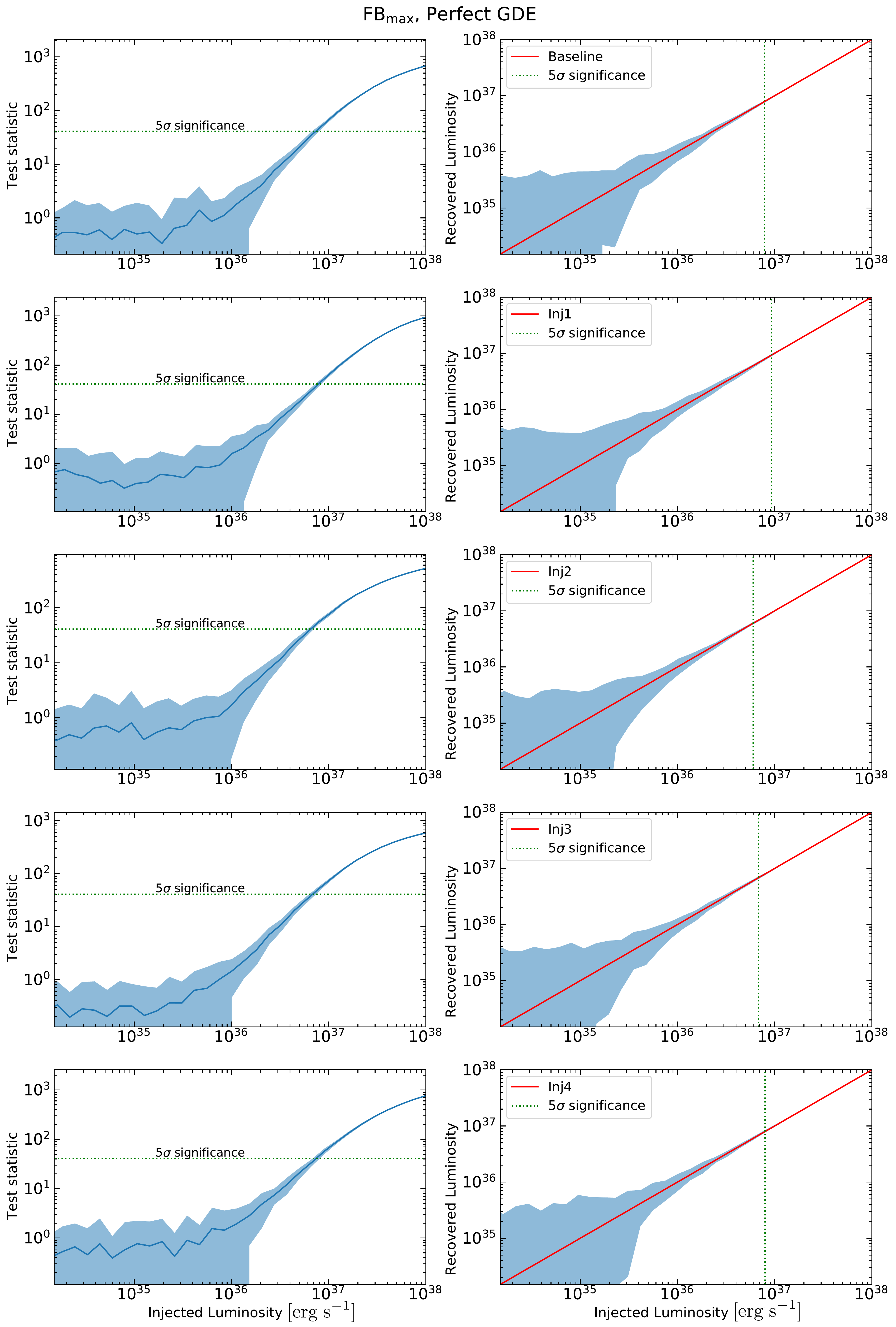}
    \caption{Same as Fig.~\ref{fig:Injection_perfectGDE}, except that here we assume the FB$_{\rm max}$ model.}\label{fig:Injection_perfectGDEFBmax}
    \end{center}
\end{figure*}

\begin{figure*}
    \begin{center}
    \includegraphics[scale = 0.48]{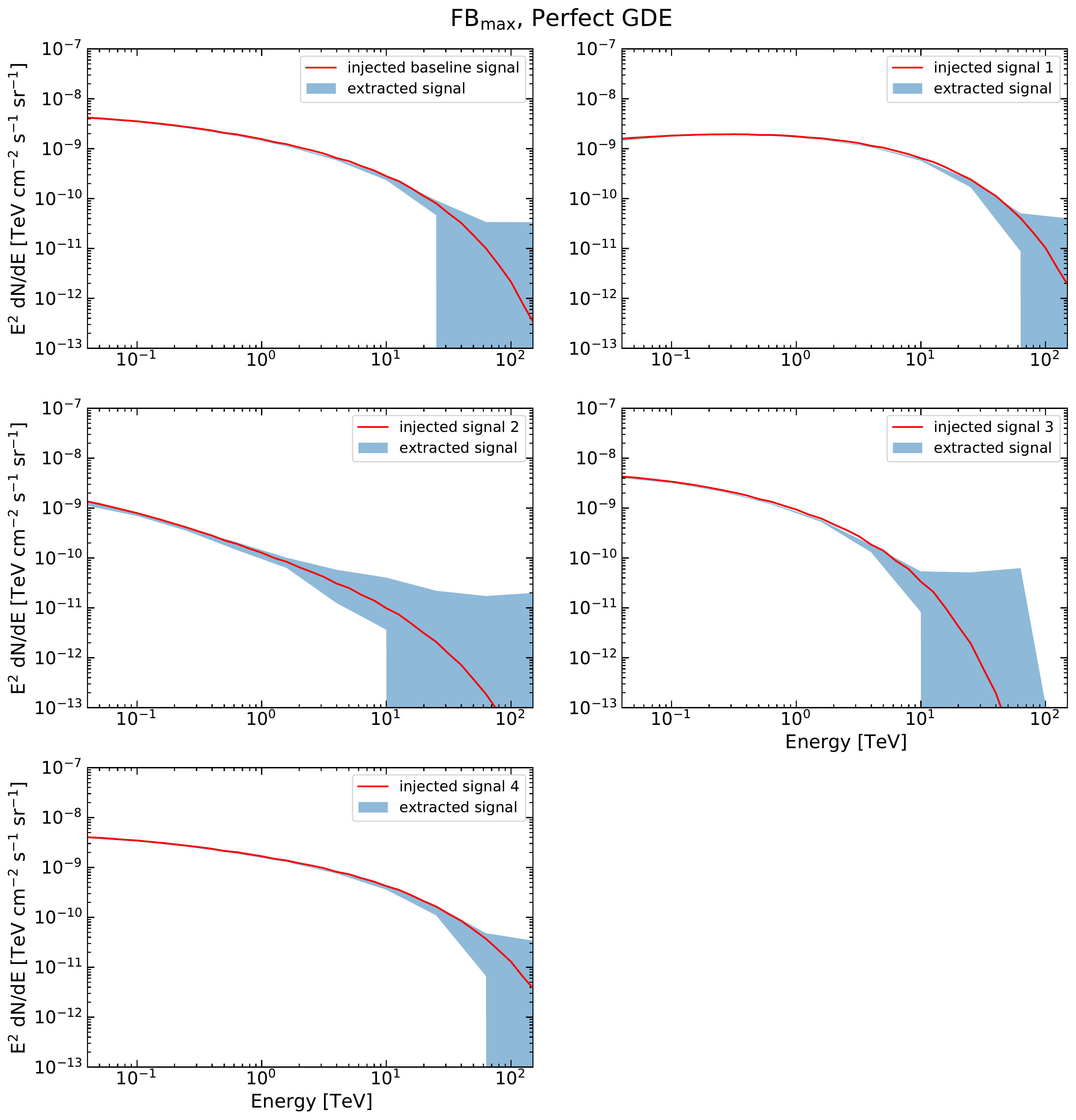}
    \caption{Same as Fig.~\ref{fig:RecoveredSpectrum_perfectGDEFBmin}, except that here we assume the FB$_{\rm max}$ model. }\label{fig:RecoveredSpectrum_perfectGDEFBmax}
    \end{center}
    \end{figure*}

\begin{figure*}
    \begin{center}
    \includegraphics[scale = 0.48]{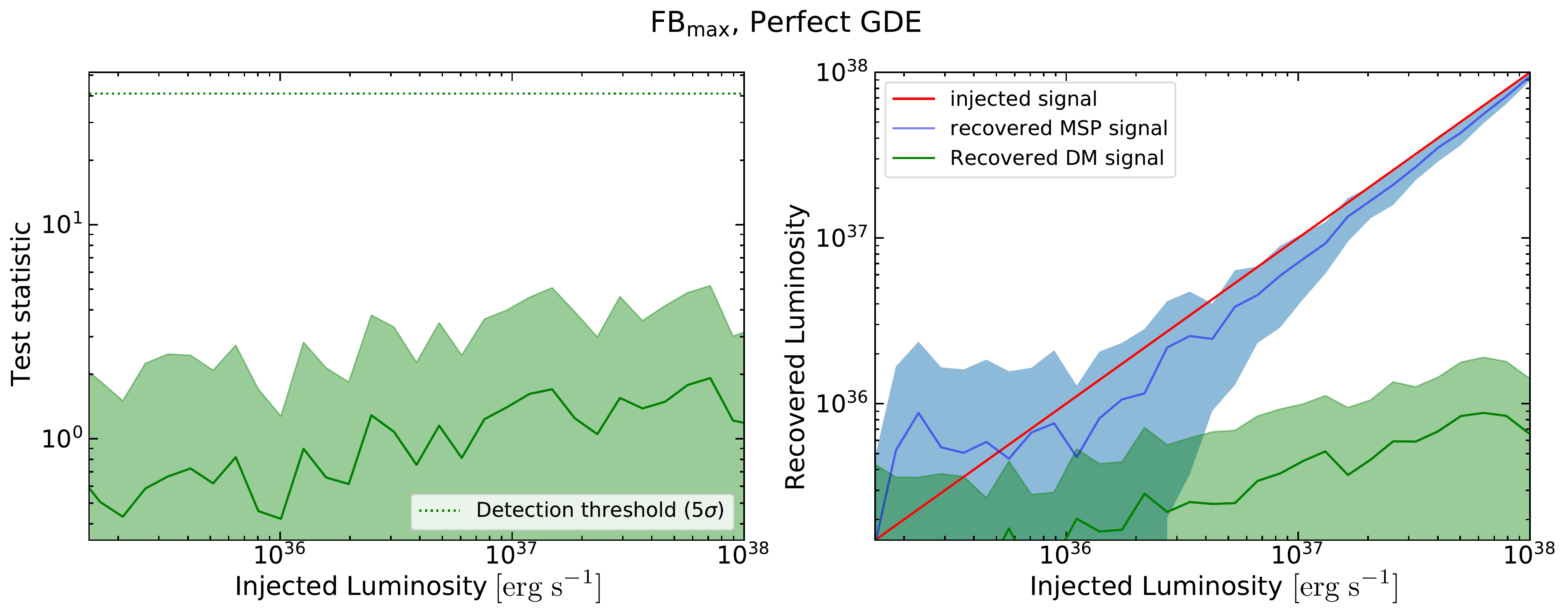}
\caption{Same as Fig.~\ref{fig:MSPsvsDM_perfectGDE}, except that here we assume the FB$_{\rm max}$ model.}\label{fig:MSPsvsDM_perfectGDEFBmax}
    \end{center}
\end{figure*}

 We have evaluated the impact of the FBs on our sensitivity by assuming the ``FB max'' model presented in Table~\ref{tab:FBmodels} and the perfect GDE model scenario. The results of this test are presented in Figs.~\ref{fig:Injection_perfectGDEFBmax}, \ref{fig:RecoveredSpectrum_perfectGDEFBmax}, and \ref{fig:MSPsvsDM_perfectGDEFBmax}; which can be directly compared Figs.~\ref{fig:Injection_perfectGDE}, \ref{fig:RecoveredSpectrum_perfectGDEFBmin}, and \ref{fig:MSPsvsDM_perfectGDE}, respectively. 

From these comparisons, it follows that the strongest impact on the CTA sensitivity to a MSPs IC signal in the GC region is due to assumptions on the FBs model. We found a degradation of the sensitivity of approximately one order of magnitude at worst---which corresponds to the MSP $e^\pm$ injection model \textit{Inj1} in Table~\ref{tab:FBmodels}---and a factor of a few for the other scenarios. As in all previous cases, we summarize the minimum luminosities for detection in Table~\ref{tab:minimumluminosities}.  

We note that in this section we have examined the degradation of the sensitivity due to uncertainties in the FBs spectrum only. However, in Appendix~\ref{appdx:FBmaxandmismodeling} we also consider the case ``FB max'' together with  mismodeling of the GDE model. Those results confirm that uncertainties in the FBs model could be the single most challenging astrophysical background component for analyses of extended $\gamma$-ray emission with CTA in the GC region.     

\section{Discussion and Conclusions}
\label{sec:discussions}

\subsection{Implied MSPs $e^\pm$ injection efficiency ($f_{e^\pm}$) for detection with CTA}
\label{sec:efficiency}
\begin{table}
  \begin{center}
    \begin{tabular}{c|c|c|c|c}
    \hline\hline  
     \rowcolor[gray]{.9}
     \multicolumn{5}{c}{ Minimum $f_{e^\pm}$ for detection [\%]}   \tabularnewline \hline
     \textbf{Baseline} & \textbf{Inj1} & \textbf{Inj2} & \textbf{Inj3} & \textbf{Inj4}\tabularnewline \hline
      \rowcolor[gray]{.9}
     \multicolumn{5}{c}{ 
     FB$_{\rm min}$, perfect GDE.}   \tabularnewline \hline
     $10.5\%$ & $2.9\%$ & $158.4\%$ & $24.3\%$ & $8.2\%$\tabularnewline \hline
      \rowcolor[gray]{.9}
     \multicolumn{5}{c}{ 
     FB$_{\rm min}$, mismodeling of the GDE.}\tabularnewline \hline
     $14.5\%$ & $3.8\%$ & $163.4\%$ & $25.3\%$ & $10.8\%$\tabularnewline \hline
      \rowcolor[gray]{.9}
     \multicolumn{5}{c}{ FB$_{\rm max}$, perfect GDE.}\tabularnewline \hline
     $57.5\%$ & $41.3\%$ & $259.4\%$ & $55.0\%$ & $58.4\%$\tabularnewline \hline
      \rowcolor[gray]{.9}\hline
     \multicolumn{5}{c}{FB$_{\rm max}$, mismodeling of the GDE.}\tabularnewline \hline
     $72.9\%$ & $51.8\%$ & $326.7\%$ & $70.4\%$ & $74.1\%$\tabularnewline 
     \hline\hline
    \end{tabular}
\caption{Minimum MSPs $e^{\pm}$ injection efficiency ($f_{e^\pm}$) required for a $5\sigma$ significance detection with CTA. The computation of these efficiencies use the implied $e^\pm$ luminosities (see Sec.~\ref{sec:efficiency}) based on the minimum luminosities reported in Table~\ref{tab:minimumluminosities}, the measured \textit{Fermi} GeV excess $\gamma$-ray luminosity, and Eq.~\ref{eq:luminosity}. The calculation of the efficiencies assumed energies greater than 700 MeV.}\label{tab:minimumfe}
  \end{center}
\end{table}

In the previous section, we  investigated the ability of CTA to characterize the TeV-scale IC $\gamma$ rays produced by an unresolved population of MSPs in the Galactic bulge region. In this section, we evaluate whether the signals that can be detected by CTA are physically possible.

 Using the relations presented in Eqs.~\ref{eq:luminosity},~\ref{eq:norm}, and assuming that the prompt $\gamma$-ray emission from the Galactic bulge population of MSPs is fully responsible for the GCE,\footnote{Note that analyses of the  
 GCE~\citep{Lacroix:2015wfx} did not detect the IC signature from the putative population of MSPs in the Galactic bulge. This might be because the prompt $\gamma$ rays are much more prominent that the IC $\gamma$ rays at GeV scale energies.}  we can obtain the minimum efficiencies $f_{e^\pm}$'s for a CTA detection of the MSPs' IC signal. In particular, we can convert the minimum IC luminosities (shown in Table~\ref{tab:minimumluminosities}) to the corresponding minimum $L_{e^\pm}$'s using Eq.~\ref{eq:norm}, and then evaluate this value in  Eq.~\ref{eq:luminosity}, along with the inferred nominal GCE luminosity [$L_{\gamma, {\rm prompt}}=2.6\times 10^{37}$ erg/s obtained in e.g.,~\citet{Macias:2019omb}]. However, the threshold IC luminosities presented in Table~\ref{tab:minimumluminosities} are estimated in the energy range 16 GeV to 158 TeV, while the GCE luminosity in~\citet{Macias:2019omb} was computed for $E_{\gamma} \gtrsim 700$ MeV. So, in order for us to connect the threshold IC luminosities with the $e^\pm$ injection luminosities that were used in our \textsc{galprop} runs, we need to extend the $e^\pm$ luminosity calculation to $700$ MeV~\footnote{This is a good approximation since the injected MSPs $e^\pm$'s can reach roughly the same minimum energies as the $\gamma$-rays. Also, note that by comparing the $e^\pm$ luminosities included in \textsc{galprop}---\textit{before propagation}---with the threshold IC luminosities, we automatically account for the effects of propagation and other energy losses (like synchrotron) for the MSPs $e^\pm$.}. 

In summary, our luminosity computations assume; $E_{e^\pm}\geq700$ MeV, $E{\gamma}\geq700$ MeV, a distance from the Sun to the GC of $8.5$ kpc (as assumed in \textsc{galprop}), and a region of interest of size $10^\circ\times10^\circ$ around the GC. It is useful to compare the fractional luminosities ($L_{e^\pm}/L_{\gamma, {\rm IC}}$) predicted by \textsc{galprop}---estimated by calculating the luminosity in the \textsc{galprop} IC maps, and the $L_{e^\pm}$ used as input in \textsc{galprop}---so as to have a better understanding of the impact of diffusion and energy losses. We obtain that $L_{e^\pm}/L_{\gamma, {\rm IC}}$ is 21.0 for the \textit{baseline} injection model, 14.3 for \textit{inj1}, 124.7 for \textit{inj2}, 23.4 for \textit{inj3}, and 21.1 for \textit{inj4} (see also Table~\ref{tab:mspspectrum}). The very large luminosity fraction obtained for \textit{inj2} is explained by the fact that this injection spectrum is very soft.       

Using the prescription described above, we are now able to compute the threshold $f_{e^\pm}$ values for the cases considered in our study. We show the results of this calculation in Table~\ref{tab:minimumfe}. Depending on assumptions about the astrophysical background components and the $e^\pm$ injection model, we obtain threshold $e^\pm$ efficiencies in the range $f_{e^\pm}\approx 2.9\%-74.1\%$, excluding \textit{inj2}---which is the softest $e^\pm$ injection spectra considered in our sample. Indeed, we obtain that the $e^\pm$ luminosity needed for CTA to detect a soft spectrum like \textit{inj2} would exceed the total budget of the MSPs spin-down energy. This means that, if the most pessimistic GDE mismodeling scenario (``FB max'' and mismodeling of the GDE) considered here is realized in nature,
CTA will only be able to reliably detect the Galactic bulge population of MSPs if the overall efficiency of this population satisfies $f_{e^\pm}\gtrsim 51.8\%$ (see the last row of Table~\ref{tab:minimumfe}). Notice that CTA might still be suited to detect this signal with percentage-level $f_{e^\pm}$'s under some specific conditions considered in Table~\ref{tab:minimumfe}.

Interestingly, the recent study of~\citet{Song:2021zrs} obtained $f_{e^\pm}\approx 10\%$, from a population analysis of the globular clusters of the Milky Way. Furthermore, dedicated models of MSP populations in globular clusters~\citep{Bednarek:2016gpp,Ndiyavala:2017hoh} have been recently constrained using MAGIC observations of the globular cluster M15. In particular, \cite{Acciari:2019ysf} constrained the electron efficiency to be $f_{e^{\pm}}\approx (0.2-2.0)\%$, for MSPs in M15. However, very likely these strong constraints cannot be directly extrapolated to other systems containing MSPs. This is because the overall apparent efficiency in globular clusters can be strongly decreased by rapid winds from Red Clump giants in globular clusters~\citep{Bednarek:2016gpp}. Note that winds can advect cosmic-ray $e^\pm$ out of the globular clusters systems before they can radiate.

Additional clues about the $f_{e^{\pm}}$ in systems containing MSP populations have been obtained in the recent analysis by~\citet{Sudoh:2020hyu}. 
Based on the break-down of a correlation 
between far-infrared and radio luminosities in SFGs, \citet{Sudoh:2020hyu} posited that radio emission from MSPs could account for a large fraction of the radio luminosity observed in systems with high stellar mass and low star formation rate. This led the authors to conclude that the MSP populations in their sample of SFGs could have $f_{e^{\pm}}$ in excess of $90\%$. They also noted, however, that several observational and theoretical uncertainties could lower their inferred efficiency to $f_{e^{\pm}}\approx 10\%$.

Given the above, 
we conclude that the efficiencies $f_{e^{\pm}}$ for MSPs are currently not very well constrained. In case that CTA makes an actual observation of the MSP IC signal in one of the scenarios disfavoured by our analysis, it could still be possible to reconcile such an observation with the MSP emission models. In particular, if the $\gamma$-ray emission from the Galactic bulge MSPs magnetosphere is beamed, only some fraction of their prompt $\gamma$-ray luminosity can be observed from the Earth. This could decrease our inferred $f_{e^{\pm}}$ by factor of a few~\citep{Sudoh:2020hyu}.  

In conclusion, we have demonstrated that---even under the assumption of high background uncertainties---if the $e^\pm$ injection spectra is harder than our \textit{inj2} model (slope of $dN/dE \propto E^{-2.5}$),  CTA has the potential to robustly discover the IC signal produced by a new population of MSPs in the GC. However, given observational and theoretical uncertainties on the $f_{e^{\pm}}$ parameter, a detection of a signal described by our \textit{inj2} model could still be possible.

\subsection{Disentangling the MSPs and DM hypotheses for the Galactic Center excess}
\label{subsec:disentanglingMSPsfromDM}

In this work, we have utilized a spatio-spectral template regression method with simulated CTA data from the GC region. In particular, we have run the signal recovery tests using a bin-by-bin analysis, which has been utilized with great success in analyses of \textit{Fermi}-LAT data [e.g.,~\citep{Ackermann:2015zua,TheFermi-LAT:2017vmf}]. This methodology allows us to reduce the impact of potential biases introduced by assumptions about the spectrum of the MSPs and/or DM templates.

Using the aforementioned method, we tested whether CTA could disentangle the IC signal produced by a unresolved population of MSPs in the Galactic bulge from the one produced by a spherical distribution of DM (or a population of pulsars following a spherical distribution). Given that the $e^\pm$ sources would follow either of these two distributions~\footnote{The reader is referred to Sec.~\ref{subsec:spatialdistribution} for details on the spatial morphologies considered in this work.}, the IC maps produced by such $e^\pm$ were predicted to have discernible spatial differences in~\cite{Song:2019nrx}. We have injected signals of MSP IC emission with varying strengths, and then run the signal recovery pipeline including both, an MSP IC template and a DM IC template in the fit. As a result, we have found that CTA has the capability of robustly disentangling these two sources, even in the presence of Galactic diffuse emission mismodeling. Overall, we have demonstrated that if CTA discovers a diffuse IC signal under the conditions considered in Table~\ref{tab:minimumluminosities}, the spatial morphology of the IC signal will reveal whether it is related to stellar mass or a spherically-symmetric source distribution as, for instance, would be expected for DM.

We note that TeV-scale $e^{\pm}$ pairs could potentially lose most of their energy very close to parent MSPs' magnetospheres. If the number of MSPs responsible for the GCE is relatively small, then the predicted IC templates could exhibit a clustering-of-photons effect~\citep{Acharyya:2020sbj}, which could facilitate the detection of the MSP population in the Galactic bulge.  Very promising methodologies to study these effects have been explored in the literature and include: the non-Poissonian template fitting procedure~\cite{Lee:2015fea,Leane:2019xiy,Chang:2019ars,Leane:2020pfc,Leane:2020nmi,Buschmann:2020adf}, wavelet techniques~\cite{Bartels:2015aea,Balaji:2018rwz,Zhong:2019ycb}, deep learning methods~\citep{Caron:2017udl,List:2020mzd}, radio detection~\citep{Calore:2015bsx,2015ApJ...805..172M,2017MNRAS.471..730R,2019ApJ...876...20H}, and X-ray detection~\citep{Berteaud:2020zef} of point sources responsible for the Galactic bulge emission.
Importantly, recent population synthesis models of MSPs~\citep{Ploeg:2020jeh} predict anywhere between $20$ and $50$ thousand MSPs in the Galactic bulge. We can use the selected spatial resolution of our simulations, and Eqs.~\ref{eq:NSD};~\ref{eq:NSC}; and~\ref{eq:boxybulge}, to obtain an estimate of the number of MSPs in each spatial 3D bin of our \textsc{galprop} simulations. We find that every bin of size $200\times200\times100$ pc$^3$ is expected to contain $\sim 300$ MSPs in the center of the boxy bulge and $\sim 30$ MSPs at 3 kpc along the long-axis of the Galactic bar. Furthermore, for the nuclear bulge region we estimate $\sim 1000$ in the central spatial bin.  This also justifies our assumption of using a smooth density function to simulate the MSPs distribution in the Galactic bulge.

\subsection{Degradation of the CTA sensitivity due to uncertainties in the astrophysical components}

 We have estimated the impact of various diffuse astrophysical components on the sensitivity to the signal. Using Monte Carlo simulations, we recreated a scenario in which the GDE is mismodeled, obtaining that if this scenario is realized in actual data, it will deteriorate the CTA sensitivity to the MSPs' IC signal luminosity at the $\lesssim 40\%$ level (depending on the characteristics of the injection spectrum and assumptions about the FBs model). 
 One possible explanation as to why this effect is not larger in our analyses, is that the intensity of the predicted GDE spectrum rapidly falls off with energy, becoming comparable to that of the expected MSP IC signal at TeV-scale $\gamma$-ray energies. We note that this is in stark contrast to analyses of \textit{Fermi}-LAT data from the GC region in which, for example, the \textit{Fermi} GeV excess signal has an intensity that is just a small fraction of the GDE emission [e.g.,~\citet{Abazajian:2020tww}].

On the other hand, we observed a drastic deterioration (up to one order of magnitude) of the IC flux  sensitivity when we switched from the ``FB min'' to the ``FB max'' model (see Table~\ref{tab:minimumluminosities}). 
The latter was proposed in~\cite{Rinchiuso:2020skh} (see the left panel of Fig.~1 in that article), as a way of conservatively accounting for the maximum $\gamma$-ray intensity that the FBs can take at the CTA energy range. This was accomplished in that work by ensuring that forthcoming measurements of diffuse $\gamma$ rays from this sky region cannot overshoot current H.E.S.S. diffuse measurements~\citep{Abramowski:2016mir} of the GC region (inner $\approx 70$ pc of the GC). As shown in Fig.~\ref{fig:counts_spectra}, the ``FB max'' model is the dominant astrophysical $\gamma$-ray component for energies greater than $\approx 30$ GeV. This explains why the FBs model produces the strongest impact on the sensitivity to the MSPs' IC signal.

Another important source of uncertainty corresponds to the spatial morphology of the FBs. The FBs template assumed in our work is a inpainted version of the spatial map in~\cite{TheFermi-LAT:2017vmf}, which in turn was constructed using a spectral component analysis using residual $\gamma$-ray data in the $1-10$ GeV range. However, the exact details of the FBs map might be susceptible to the energy range that is included in the spectral component analysis of~\cite{TheFermi-LAT:2017vmf}. In a future study, we will address the impact of spatial uncertainties in the FBs model, on the CTA sensitivity to an MSP IC signal in the GC.  

\subsection{Impact of the MSPs' $e^\pm$ injection spectrum on the sensitivity to the IC signal}

As discussed in the previous section, one of the most important factors determining the characteristics of the expected IC signal, corresponds to the $e^\pm$ injection spectrum [see also~\citep{Song:2019nrx}]. In our analysis we  considered five different $e^\pm$ injection spectra, finding that these produce significantly different results. For example, the injection model \textit{Inj1} (see Table~\ref{tab:minimumluminosities}) has a detection threshold luminosity that is a factor of $\sim 4$
lower than that of the injection model \textit{Inj2} in both ``FB min'' cases under consideration. The most likely explanation for this is that signals that mirror the spectral shape of the GDE---see the left-middle panel of Fig.~\ref{fig:RecoveredSpectrum_perfectGDEFBmin} in comparison to the left panel of Fig.~\ref{fig:counts_spectra}---are more difficult
to disentangle than the signals that have a distinct spectral shape.

In Sec.~\ref{subsec:e+-injection} we presented details about our assumptions on the injection spectra of $e^\pm$ pairs from MSPs. Although each individual MSP---that makes up the GCE---very likely has different properties (e.g., spin-down luminosity, age, and stellar surface magnetic field), in this work we have computed the expected IC emission from MSPs by assuming a mean $e^\pm$ injection spectra for the whole MSPs population. This assumption is motivated by studies such as that by \citet{Kalapotharakos:2019cio} in which, using \textit{Fermi}-LAT data of resolved MSPs and young pulsars, it was determined that there is a correlation between MSP gamma-ray luminosity, spin-down power, spectral energy cutoff, and stellar surface magnetic field strength.  

We note that in our study, the normalization of the $e^\pm$ injection spectrum is fixed by our Eq.~\ref{eq:norm}, such that prompt gamma-ray emission from the MSPs population explains the GCE data. If the MSPs population happened to have a rather low mean spin-down luminosity, a larger number of MSPs would be required to explain the GCE, and so our predictions would possibly not be affected. However, a lower spin-down luminosity might likely also imply a lower energy cutoff $E_{\rm cut}$ in Eq.~\ref{eq:mspspectrum}. We have investigated the effect of the $E_{\rm cut}$ with our \textit{Inj3} and \textit{Inj4} models. As can be seen in Tab.~\ref{tab:minimumluminosities} and \ref{tab:minimumfe}, it becomes more difficult to detect the IC signal from the bulge MSPs population as $E_{\rm cut}$ decreases. Other MSPs parameters---such as pulsar age, and stellar surface magnetic field---are expected to also affect the spectral slope, and the energy cutoff of the $e^\pm$ injection spectrum. The range of parameters evaluated in this work cover a generous range of possible parameters to account for these theoretical uncertainties in MSPs modeling.

The modelling of the distribution of MSPs $e^\pm$ throughout the Galactic bulge is based on a steady-state assumption, where the spatial distribution of MSPs is described using a smoothly varying function of position that does not change with time and the MSPs are assumed to inject $e^\pm$ at a constant rate. While in reality MSPs must have a proper motion and a certain decay time, our assumption of a continuous emission scenario is an approximations that is justified given the potentially low proper motion of MSPs~\citep{Hobbs:2005yx} and small period derivatives ($\dot{P}$)~\citep{Harding:2021yuv}. We leave for future work to explore the effect that a fully time-dependent modeling of MSPs would have in our results, but we anticipate a potentially small effect given the aforementioned arguments.

Future advances in our understanding of the MSPs' $e^\pm$ injection spectrum will help reduce the impact of this component on the sensitivity. On the theory side, future global magnetosphere simulations that include non-dipolar fields could provide a clearer picture of the $e^\pm$ injection mechanisms and predicted spectrum~\citep{Harding:2021yuv}. On the phenomenological side, multiwavelength measurements and modeling of unresolved MSPs in globular clusters~\citep{Ndiyavala:2017hoh, Song:2021zrs} and SFGs~\citep{Sudoh:2020hyu}, could reveal the characteristics of the MSPs $e^\pm$ injection spectra. 

\section*{Acknowledgements}

We thank Mar\'ia Benito, Christopher Eckner, Fabio Iocco,  Anastasiia Sokolenko, Gabrijela Zaharijas, and the CTA dark matter working group for useful discussions. We also thank the anonymous referee for useful comments that have improved the manuscript. O.M. acknowledges support by JSPS KAKENHI Grant Number JP20K14463.  The work of D.S.\ is supported by the U.S.\ Department of Energy under the award number DE-SC0020262.
S.A. is supported by JSPS/MEXT KAKENHI Grant numbers JP17H04836 and JP18H04340.
The work of S.H.\ is supported by the U.S.\ Department of Energy under the award number DE-SC0020262 and NSF Grant numbers AST-1908960 and PHY-1914409. This work was supported by World Premier International Research Center Initiative (WPI Initiative), MEXT, Japan. R.M.C. acknowledges support from the Australian Government through the Australian Research Council for grant DP190101258 shared with Prof.~Mark Krumholz at the ANU. D.M.N. acknowledges support from NASA under award Number 80NSSC19K0589. The authors acknowledge Advanced Research Computing at Virginia Tech for providing computational resources and technical support that have contributed to the results reported within this paper.
\section*{data availability}
The astrophysical templates underlying this article will be shared on reasonable request to the corresponding author.




\bibliographystyle{mnras}
\bibliography{references} 


\appendix

\section{Stellar density profiles}
\label{appdx:Stellardensity}

\subsection{Nuclear bulge}
Motivated by recent reanalysis of the GCE [e.g.,~\cite{Macias:2019omb}], we assumed here that the MSP population that accounts for the GCE follows the distribution of stars in the nuclear bulge and the boxy bulge. For the nuclear bulge, we used the three-dimensional density function given in~\cite{Launhardt:2002tx} [see also~\cite{Song:2019nrx}]. As mentioned in Sec.~\ref{sec:backgroundtemplates}, the nuclear bulge consists of the nuclear stellar disk (NSD) and the nuclear stellar cluster (NSC). The NSD density function is given by
\begin{equation}\label{eq:NSD}
\rho_{\rm NSD}(x, y, z)=\left\{\begin{array}{ll}\rho_{0}\left(\frac{r}{r_{d}}\right)^{-0.1} \exp \left(-\frac{|z|}{z_{d}}\right), & \textrm{for}\; r / \mathrm{pc}<120 \\ \rho_{1}\left(\frac{r}{r_{d}}\right)^{-3.5} \exp \left(-\frac{|z|}{z_{d}}\right), & \textrm{ for } 120 \leq r / \mathrm{pc}<220 \\ \rho_{2}\left(\frac{r}{r_{d}}\right)^{-10} \exp \left(-\frac{|z|}{z_{d}}\right), & \textrm{for}\; r / \mathrm{pc} \geq 220 \end{array}\right.    
\end{equation} 
where $r^2=x^2+y^2+z^2$, $r_{d}=1$ pc, $z_{d}=45$ pc and $\rho_{0}=301 \mathrm{M}_{\odot}$ pc$^{-3}$. The total stellar mass of the NSD in the inner $120$ pc is set to $8 \times 10^{8}\;\mathrm{M}_{\odot}$, which then defines $\rho_{1}$, and $\rho_{2}$ through continuity conditions. 

In addition, the NSC is given by
\begin{equation}\label{eq:NSC}
    \rho_{N S C}(x, y, z)=\left\{\begin{array}{ll}\rho_{3}\left[1+\left(\frac{r}{r_{c}}\right)^{2}\right]^{-1}, & \textrm{for}\; r / \mathrm{pc}<6 \\ \rho_{4}\left[1+\left(\frac{r}{r_{c}}\right)^{3}\right]^{-1}, & \textrm{for}\; 6<r / \mathrm{pc} \leq 200 \\ 0, & \text { for } r / \mathrm{pc}>200\end{array}\right.
\end{equation}
where $r_c= 0.22$ pc, and  $\rho_3 = 3.3 \times 10^6 \; \mathrm{M}_{\odot}$. The density $\rho_4$ is obtained by imposing continuity at $r=6$ pc.

\subsection{The boxy bulge}

The boxy bulge stellar density function was derived by~\cite{Freudenreich:1998}. We used the Model S in that reference, which is given by

\begin{equation}
\rho_{\mathrm{bar}}(R, \phi, z) \propto  \operatorname{sech}^{2}\left(R_{s}\right)
 \times\left\{
 \begin{array}{lr}
1, & R \leq R_{\mathrm{end}} \\
e^{-\left[\left(R-R_{\mathrm{end}}\right) / h_{\mathrm{end}}\right]^{2}}, & R>R_{\mathrm{end}}
\end{array}\right.\label{eq:boxybulge}
\end{equation}
where $R_{\mathrm{end}}=3.128$ kpc, $h_{\mathrm{end}}=0.461$ kpc, and $R_s$ is defined as
\begin{equation}
\begin{aligned}
R_{\perp}^{C_{\perp}} &=\left(\left|X^{\prime}\right| / a_{x}\right)^{C_{\perp}}+\left(\left|Y^{\prime}\right| / a_{y}\right)^{C_{\perp}} \\
R_{s}^{C_{\|}} &=R_{\perp}^{C_{\mid 1}}+\left(\left|Z^{\prime}\right| / a_{z}\right)^{C_{\|},}
\end{aligned}    
\end{equation}
with  $a_{x}=1.696$ kpc, $a_{y}=0.6426$ kpc, $a_{z}=0.4425$ kpc, $C_{\|}=3.501$, and $C_{\perp}=1.574$.  Here, the coordinates $(X^\prime,Y^\prime,Z^\prime)$ correspond to a Cartesian coordinate system in the boxy bulge frame of reference (see~\citealt{Song:2019nrx} for more details).

\subsection{Spherically symmetric template}

We used a generalized NFW density profile of the form 
\begin{equation}\label{eq:NFW}
    \rho_{\rm NFW}(r)=\frac{\rho_{0}}{\left(\frac{r}{R_{\odot}}\right)^{\gamma}\left(\frac{1+r / R_{s}}{1+R_{s} / R_{\odot}}\right)^{(3-\gamma)}}
\end{equation}

where $\alpha=1$, $\beta=3$, $R_{\odot}=20$ kpc, and $\gamma=1.2$. Previous studies of the GCE~\citep{Abazajian:2012pn,Gordon:2013vta,Daylan:2014rsa,Calore:2015bsx} showed that the square of this density profile ($\rho^2_{\rm NFW}$)  was a good fit to the signal morphology. Although more recent analyses~\citep{Macias:2016nev,Bartels:2017vsx,Macias:2019omb,Abazajian:2020tww} using improved background models have shown that stellar mass profiles of the Galactic bulge give a much better fit to the data, it is still interesting to consider this morphology because it  would test the ability of CTA to detect the IC signature from DM self-annihilation.

\section{Impact of the Fermi bubbles model in the presence of Galactic diffuse emission mismodeling}
\label{appdx:FBmaxandmismodeling}

\begin{figure*}
    \begin{center}
    \includegraphics[scale = 0.42]{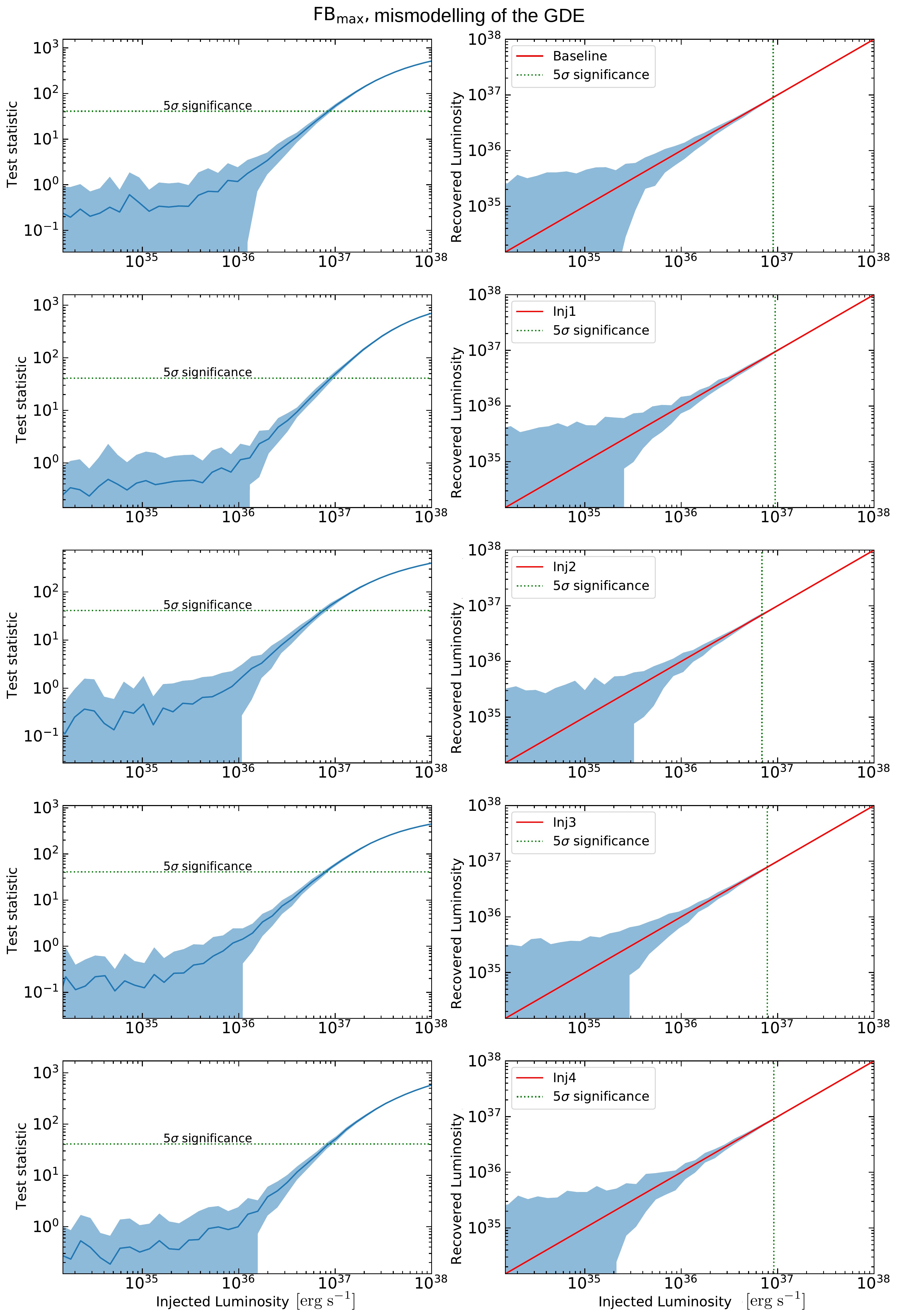}
    \caption{Same as Fig.~\ref{fig:Injection_perfectGDE}, except that here we assume mismodeling of the Galactic diffuse emission model and the FB$_{\rm max}$ model.}\label{fig:InjectionmismodelingFBmax}
    \end{center}
\end{figure*}

\begin{figure*}
    \begin{center}
    \includegraphics[scale = 0.48]{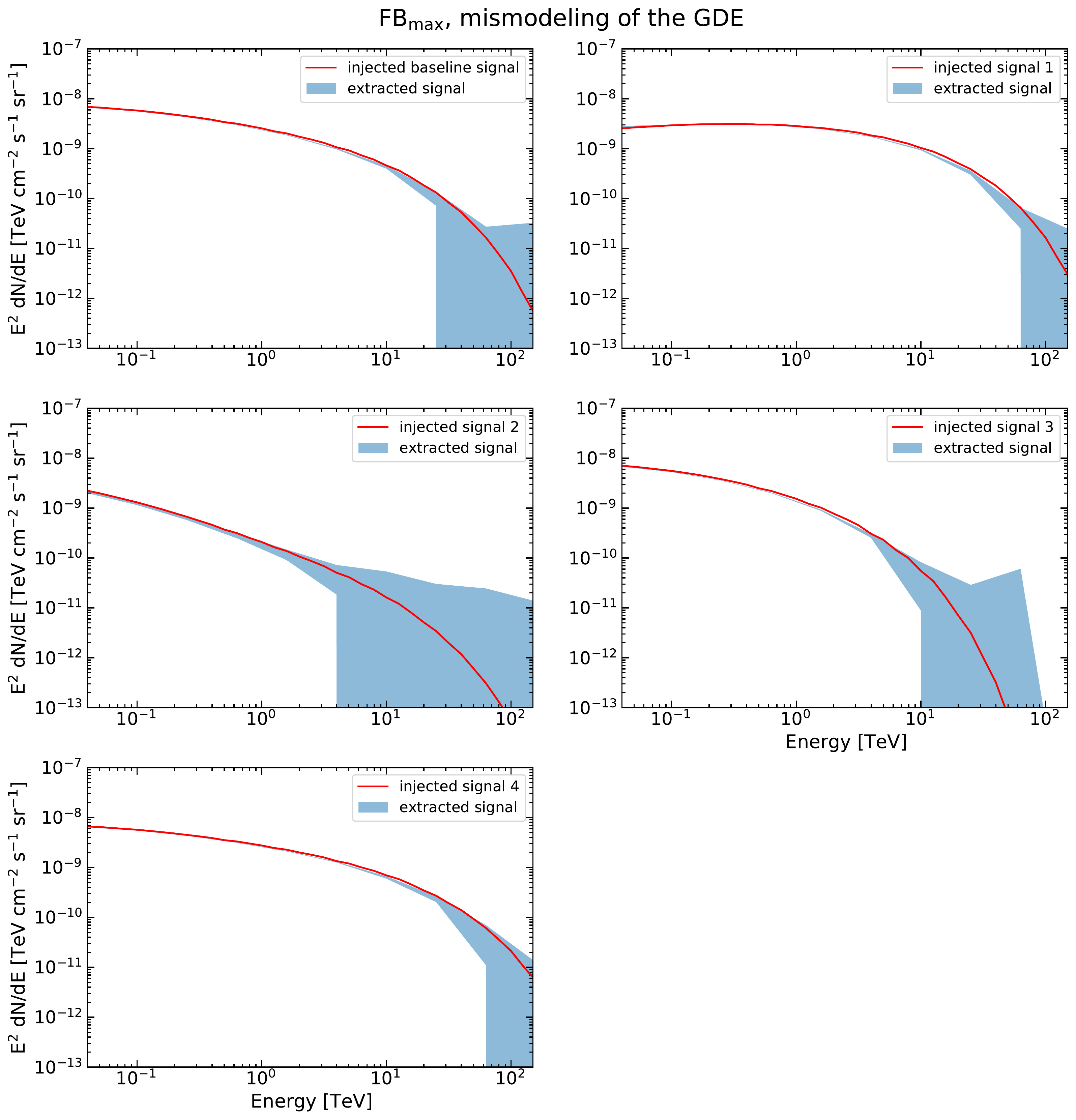}
    \caption{Same as Fig.~\ref{fig:RecoveredSpectrummismodelingGDEFBmin}, except that here we assume the FB$_{\rm max}$ model. }\label{fig:RecoveredSpectrum_mismodelingGDEFBmax}
    \end{center}
    \end{figure*}

\begin{figure*}
    \begin{center}
    \includegraphics[scale = 0.48]{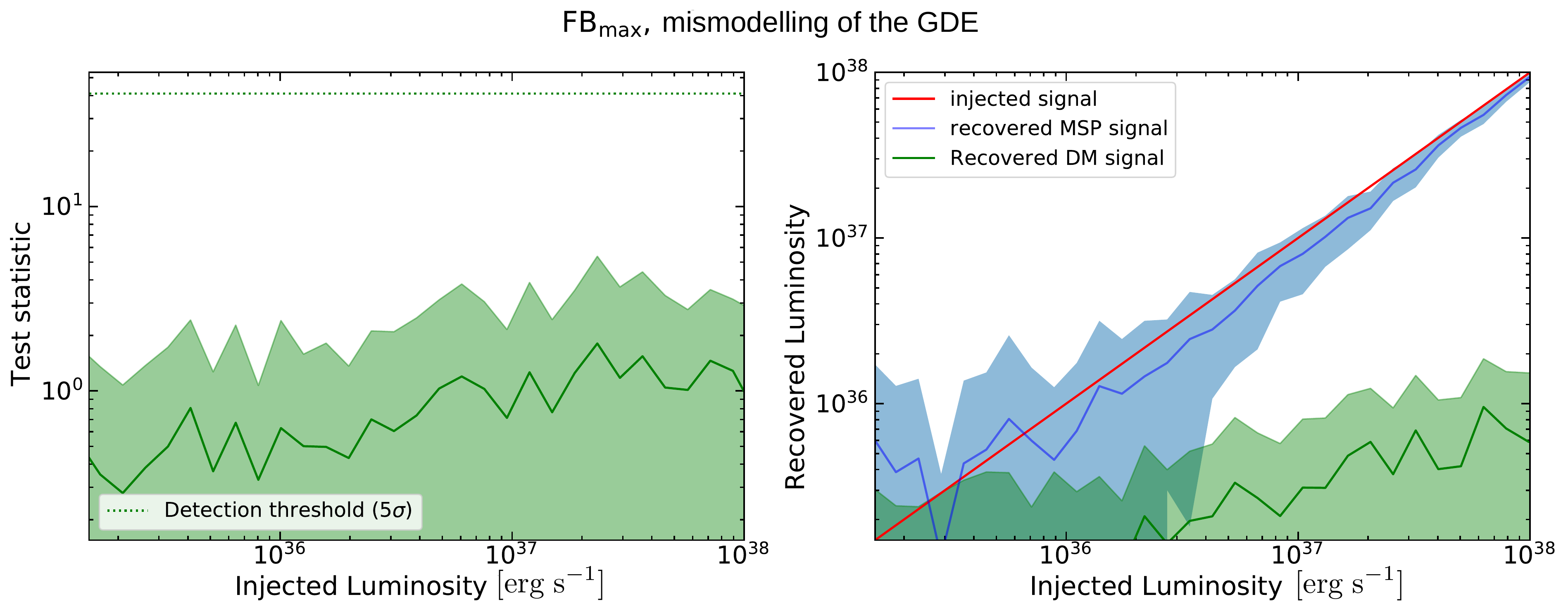}
    \caption{Same as Fig.~\ref{fig:MSPsvsDM_perfectGDE}, except that here we assume mismodeling of the Galactic diffuse emission model and the FB$_{\rm max}$ model.}\label{fig:MSPsvsDM_mismodelingGDEFBmax}
    \end{center}
    
\end{figure*}

Here, we present the results of the systematic uncertainties analysis in the case where the FBs component have the maximum possible intensity (FB$_{\rm max}$ scenario) and the GDE components are mismodeled. We show the signal recovering tests in Fig.~\ref{fig:InjectionmismodelingFBmax}, the minimum flux for detection in Fig.~\ref{fig:RecoveredSpectrum_mismodelingGDEFBmax}, and the MSPs vs. DM degeneracy tests in Fig.~\ref{fig:MSPsvsDM_mismodelingGDEFBmax}. We also present the threshold IC luminosities and $f_{e^\pm}$ in Table~\ref{tab:minimumluminosities}, and Table~\ref{tab:minimumfe}, respectively. 

 We note that this constitutes the most challenging scenario for a reliable CTA detection of the MSPs IC signal. As was shown in the left panel of Fig.~\ref{fig:counts_spectra}, our FB$_{\rm max}$ model outshines all other components at the CTA energy range. We remind the reader that this model was constructed by imposing that the FBs model does not overshoot the H.E.S.S. diffuse  measurements~\citep{Abramowski:2016mir} from the inner $\approx 70$ pc of the GC. Tough the exact spatial morphology and spectra of the low-latitude part of the FBs are very uncertain, our very conservative modeling of this component allows to show that CTA has the capability to reliably discover the IC signature from an unresolved population of MSPs in the Galactic bulge.    
 
 \section{Comparisons with previous results in the literature}
 
 In~\citet{Hooper:2018fih}, it was pointed out that if MSPs are efficient ($f_{e^\pm}\approx 10\%$) accelerators of multi-TeV $e^\pm$, then TeV and radio
synchrotron emission from these sources could be used to constrain the population of MSPs present in the inner Milky Way. In particular, in that paper it was shown that if MSPs are responsible for the GCE, then these are expected to saturate or overshoot the TeV emission observed from the inner $\approx 0.2^\circ-0.5^\circ$~\citep{Abramowski:2016mir}. 
 
 We note that a direct comparison of our results with those in \citet{Hooper:2018fih} is at present not possible. This is due to the fact that the spatial resolution of our \textsc{galprop v56} simulations (see Tab.~\ref{tab:galpropsetup}) are not sufficient to probe the inner $\approx 0.2^\circ-0.5^\circ$ of the Galaxy. However, it should also be noticed that in our pipeline we have masked the Galactic plane region defined by $|b|<0.3^\circ$. In Fig.1 (left) of \citep{Rinchiuso:2020skh}, some of us have shown that the HESS Pevatron spectra~\citep{Abramowski:2016mir} averaged over our $10^\circ \times 10^\circ$ RoI are higher than the predicted fluxes that are observable by CTA (see Tab.~\ref{tab:minimumluminosities}).

Even though the Galactic ridge region~\citep{Aharonian:2006au} is excluded from our analysis, it is useful to study the predictions of our model for this sky region. We show this in Fig.4 of ~\citep{Song:2019nrx}. As can be seen in that figure,
we obtained that for $f_{e^\pm} = 10\%$, our hard injection spectrum (\textit{Inj2}) starts to be in tension with the H.E.S.S. measurements. However, \citet{Song:2019nrx} also investigated how to reduce this apparent tension. Though we demonstrated that our model predictions are not very sensitive to changes in the propagation parameters$-$if these are taken to be within the 95\% credible contours provided by~\citet{Johannesson:2016rlh}. We found that the tensions with the H.E.S.S. measurements can be relaxed if we assume stronger magnetic fields than the ones assumed in the default GALPROP setup. In particular, Fig.~8 of \citet{Song:2019nrx} shows the impact of changing the magnetic field from  $\approx 10\; \mu$G to $50\;\mu$G~\citep{Crocker:2010xc} in the nuclear bulge region. As can be seen, this change reduces the IC emission in the Galactic ridge by a factor of $1/2$.  Here, it should be noticed that for this test we used a constant B-field of 50 $\mu$G in the inner 400 pc of the Galaxy, but outside this region it matches the default exponential B-field model in \textsc{galprop v56}.  

\begin{figure}
    \begin{center}
    \includegraphics[scale =1.0]{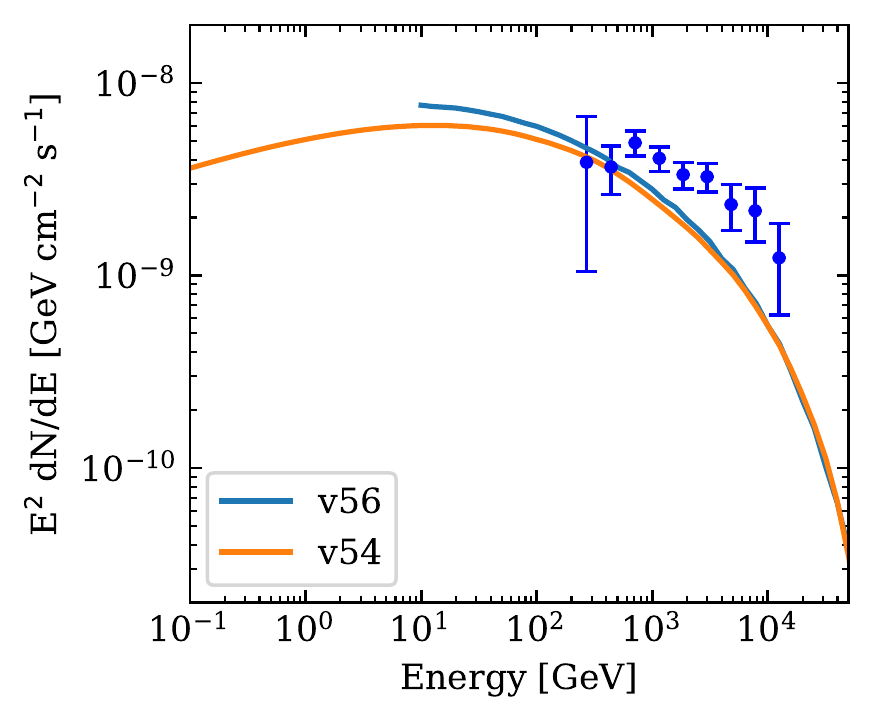}
    \caption{Comparison between the $\gamma$-ray spectra predicted with \textsc{galprop v54} and \textsc{galprop v56}. The spectrum assumes the inner $6^\circ \times 2^\circ$ of the GC, the default magnetic field of 10 $\mu$G at the GC, $f_{e^\pm}=10\%$, and the data points are the H.E.S.S.~\citep{Aharonian:2006au} measurements of the Galactic ridge region.   }\label{fig:compareV54andV56}
    \end{center}
    
\end{figure}

As mentioned in Sec.~\ref{sec:ICfromMSPs}, an important difference between the present work and the one in \citep{Song:2019nrx}, is that in the former we used \textsc{galprop v56}, while in the later we used \textsc{galprop v54}. Given that there are substantial improvements in the new version of the code, it is important to estimate the differences between these two versions. In Fig.~\ref{fig:compareV54andV56} we present a comparison between our baseline injection model computed with V54 and V56. Following ~\citep{Song:2019nrx}, we show the comparison for the Galactic ridge region, and assuming the default 10 $\mu$G at the GC position. As can be seen, we obtain small differences between the normalization of the IC predictions using the two \textsc{galprop} versions.

 \section{Propagation of Galactic bulge MSPs positrons to Earth}
 
\begin{figure}
    \begin{center}
    \includegraphics[scale =1]{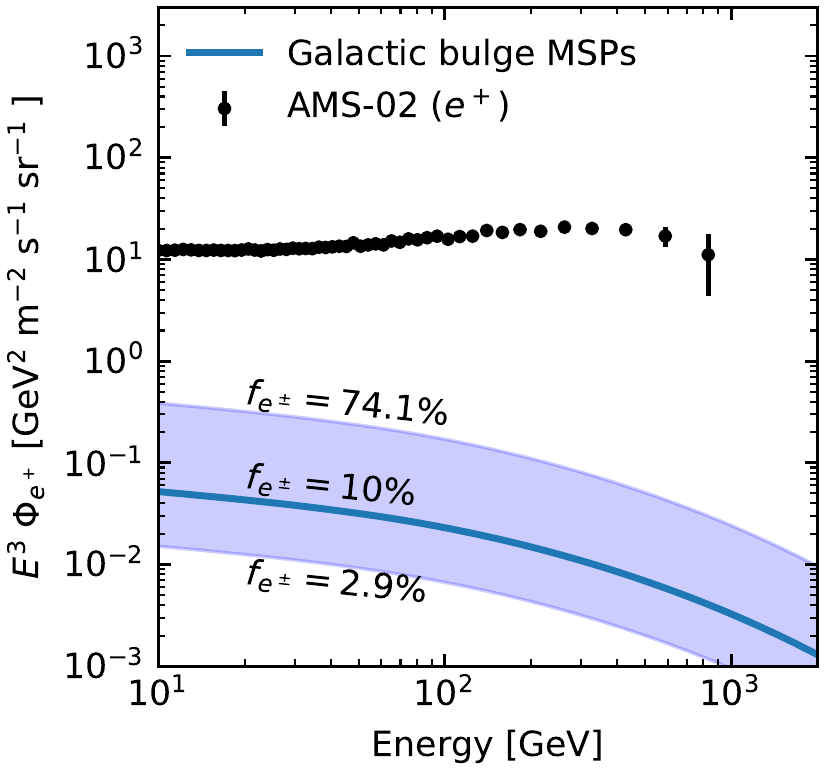}
    \caption{Total Galactic bulge MSP contribution (assumed to be equal numbers of positrons and electrons) to the measured CRs $e^+$ spectrum at Earth~\citep{Aguilar:2021tos}. The blue band shows the $e^\pm$ injection efficiencies in the range $\approx 2.9$--74.1\%, which are required for a robust CTA detection of the IC emission from the Galactic bulge population that explains the GCE. The $e^\pm$ injection spectra corresponds to the baseline in Table~\ref{tab:mspspectrum}.  The positrons from Galactic bulge MSPs have been propagated using \textsc{GALPROP} v56 and the propagation parameter setup in Table~\ref{tab:galpropsetup}. We have neglected the effect of solar modulation for energies greater than 10 GeV, as suggested by~\citet{Strauss:2015}.}\label{fig:MSPpositronsatEarth}
    \end{center}
    
\end{figure}

We have propagated the $e^\pm$ pairs---injected by the putative GC population of MSPs---till Earth, and compared the predicted $e^+$ fluxes (assuming equal number of $e^+$ and $e^-$ injected by MSPs) with recent measurements by the AMS-02 telescope~\citep{Aguilar:2021tos}. Figure~\ref{fig:MSPpositronsatEarth} displays the results of our simulations for the baseline injection scenario (see Table~\ref{tab:mspspectrum}) and \textsc{galprop} propagation setup (see Table~\ref{tab:galpropsetup}). The blue band shows the range of efficiencies ($f_{e^\pm}\approx 2.9-74.1\%$) that would be required for a robust CTA detection of the IC signal under different levels of Galactic diffuse emission mismodeling. Our results indicate that the predicted MSPs contribution to the observed AMS-02 events is very small, and difficult to detect, even for very high $f_{e^\pm}$. This is not surprising given that $e^\pm$ lose energy very efficiently through Synchrotron and IC radiation.

A number of recent studies~\citep[e.g.,][]{Cholis:2018izys, Manconi:2020ipm} have investigated the contribution of pulsars to the AMS-02 measurements. Such works suggest that even for relatively small $f_{e^\pm}$ values, local pulsars could explain the AMS-02 observations. So, one might wonder whether such results are in tension with the required GC MSPs efficiencies for a CTA detection. As shown in Fig.~\ref{fig:MSPpositronsatEarth}, it would be very difficult to detect $e^\pm$ accelerated by bulge MSPs at Earth. So, in order to constrain the $f_{e^\pm}$ through this method, one would need to assume that $f_{e^\pm}$ of the bulge MSPs are the same as those of the field pulsars nearer Earth. However, the local pulsars are mostly young to middle-aged~\citep{Delahaye:2010}, and these could have systematically different $e^\pm$ efficiencies than the more evolved Galactic bulge MSPs population~\citep{Song:2021zrs}. Though a detailed quantitative analysis of the impact of the local pulsars in the AMS-02 is beyond the scope of this paper, it should be noted that it is difficult to compare our predicted efficiencies with those in~\citep[e.g.,][]{Cholis:2018izys, Manconi:2020ipm} because these are highly dependent of environmental parameters such as magnetic fields, diffusion coefficients, pulsar distance measurements, etc.    

\bsp	
\label{lastpage}
\end{document}